\documentclass[prd,aps,floatfix,nofootinbib,preprint,tightenlines]{revtex4}
\usepackage{dcolumn}
\usepackage{amsmath}
\usepackage{latexsym}
\usepackage{graphicx}
\usepackage{bm}

\long \def \blockcomment #1\endcomment{}

\newcommand{\bee}{\begin{equation}}
\newcommand{\ee}{\end{equation}}
\newcommand{\beea}{\begin{eqnarray}}
\newcommand{\eea}{\end{eqnarray}}

\begin{document}
\title{
Lattice calculations of the spectroscopy of baryons with broken flavor $SU(3)$ symmetry and 3, 5, or 7 colors}
\author{Thomas DeGrand}
 \affiliation{Department of Physics,
University of Colorado, Boulder, CO 80309, USA}

\begin{abstract}
Lattice Monte Carlo calculations of baryon spectroscopy in gauge groups $SU(N)$, $N=3$, 5, 7,
are presented. The quenched valence fermions come in three flavors,
two degenerate mass ones and a third heavier flavor. The data shows striking regularities
reminiscent of the real-world case of $N=3$: higher angular momentum states lie higher in mass, 
and Sigma-like states lie higher than Lambda-like ones. These simple regularities 
are reasonably well described by $1/N$ expansions.
\end{abstract}

\maketitle

\section{Introduction}
QCD in the the limit of a large number of colors has a long history as a tool for the study of the
strong interactions, going back to the initial studies of Refs.~\cite{'tHooft:1973jz,'tHooft:1974hx}.
Much of this work is done in the continuum, but there is also a substantial body of lattice studies
of large-$N$ gauge theories. The subject has recently been reviewed in Ref.~\cite{Lucini:2012gg}.

Baryon spectroscopy represents an interesting corner of large-$N$ investigation.
Baryons in large $N$ can be analyzed as many-quark states \cite{Witten:1979kh}
 or as topological objects
in effective theories of mesons\cite{Witten:1983tx,Adkins:1983ya}.
Large-$N$ mass formulas for baryons have been developed by the authors of 
Refs.~\cite{ Jenkins:1993zu,Dashen:1993jt,Dashen:1994qi,Jenkins:1995td,Dai:1995zg,Cherman:2012eg},
and work up to 1998 has been summarized in a review, Ref.~\cite{Manohar:1998xv}.

Until recently, comparisons of  large-$N$ expectations to data have only been possible for $N=3$, either
 to real-world baryon spectroscopy,
or to lattice Monte Carlo data for $N=3$ baryons \cite{Jenkins:2009wv}. The latter case makes an important
contribution to large-$N$ phenomenology, because large-$N$ predictions are not restricted to physical quark masses;
they should be applicable across a wide range of quark masses.
(Of course, the chiral limit, within the context of large-$N$ QCD, is itself an interesting problem.)

Comparing large-$N$ predictions with results from several $N$'s has an obvious advantage over making
comparisons only to $N=3$. In particular, it allows one to directly observe the approach to some limiting
 behavior from $N=3$.
Recently, I  presented the first lattice simulation results on baryon spectroscopy which were directly
appropriate to
large-$N$ QCD, by comparing baryons made of two flavors of degenerate mass quarks 
and $N=3$, 5, and 7 colors \cite{DeGrand:2012hd}. The calculations used the quenched approximation for simplicity. 
The qualitative expectations of large-$N$ phenomenology
(to be recapitulated below) were observed in the spectroscopy.
The present paper is a modest extension of that work, to look at the spectroscopy of baryons with three flavors
of valence quarks, mimicking the real world with, again, a mass-degenerate pair of nonstrange quarks
 and a third heavier strange quark. Striking regularities are observed in the patterns of mass splittings,
regularities which are consistent with simple parametrizations motivated either by the quark model or (more generally)
by large-$N$ counting.

By modern standards for simulation of $N=3$ QCD, these simulations are extremely naive. 
They involve a single lattice spacing and a single simulation volume. More importantly, they employ
the quenched approximation. This approximation is uncontrolled, and so is no longer used in $N=3$ QCD.
However, away from the deep chiral limit, it is hard to see the effects of un-quenching on the hadron spectrum;
the qualitative features already expected from quark model calculations were seen in the earliest quenched lattice
 studies of spectroscopy. One can argue that the quenched approximation becomes ever better motivated
as $N$ becomes large. I used it, though, simply because it was cheap and because it was known not to
produce wildly inaccurate results for $N=3$.

Interested researchers could, if they wished, eliminate all these shortcomings
 in the same way that they have been
 gradually eliminated
in ordinary QCD simulations, by simulating at several volumes and several lattice
 spacings, and by incorporating
dynamical fermions. The cost would probably scale like $N^3$, compared to a usual
QCD simulation, from the cost of multiplying $SU(N)$ matrices. But, it
 is worth pointing out (this was first stated by the authors of
 Ref.~\cite{Jenkins:2009wv}): ``An important
observation is that the $1/N$ counting rules hold at finite lattice spacing, 
and so are respected by the lattice results
$\it including$ the finite lattice spacing corrections dependent on the lattice spacing $a$.''
Thus there appears to be ample justification for beginning as simply as possible.

My presentation of results will be given in the context of models or of expressions derived from large-$N$ counting.
Lattice calculations of spectroscopy or matrix elements in $N=3$ are usually not shown this way.
 The reason for my approach is that the question associated with the large-$N$ limit has always
 been ``to what extent does $N=3$ spectroscopy match large-$N$ expectations?'' Answering that question involves
comparing data to some theoretical construct.

People who are more focused on large-$N$ phenomenology than on lattice simulation should be aware of two things:

First, $1/N$ analyses assume that there is a hierarchy of interaction strengths, and that
 $1/N$ is actually a small number. For example,
$N$-color baryons with flavor $SU(2)$ symmetry come in isospin-spin locked multiplets, with isospin $I$
 and angular momentum $J$ equal to
$I=J=1/2$, $3/2,\dots,N/2$. The spectrum is rotor-like, $M(N,J)\sim N m_0 + BJ(J+1)/N + \dots$. If the second term
is to be small, then one needs $J\ll N$. However, a lattice simulation can compute the mass of a baryon with any
of the mentioned $J$ values, and, it turns out that it is much less expensive to construct the correlation function for
a large-$J$ state than for a small-$J$ state. So the analysis will make extensive use of large-$J$ states.
 One might worry that this would 
invalidate a large-$N$ approach. However, it  will happen
that the spectroscopy of all states (at least for the $N$'s I studied) is well described by large-$N$ counting.
I believe that this happens simply because $B$ is smaller than $m_0$.

Second, quark models or large-$N$ analyses typically give mass formulas for baryons of color $N$, 
angular momentum $J$, isospin $I$, number of strange quarks $N_s$ and nonstrange and strange quark masses
$m_{ns}$ and $m_s$, $M(I,J,N,N_s,m_{ns},m_s)$, in terms of a set of coefficients of particular functions of
the dependent variables. Often, these coefficients are given in terms of explicit combinations of the masses
of particular states. For example, with two degenerate flavors, and the rotor formula
 \bee
 M(N,J) = Nm_0  + \frac{J(J+1)}{N}B ,
\ee
one can extract the coefficients from two-particle differences
\bee
m_0=\frac{5}{4N}M(N,J=1/2) - \frac{1}{4N}M(N,J=3/2),
\label{eq:botA}
\ee
or
\bee
m_0=\frac{N+2}{4N}M(N,J=N/2-1) - \frac{N-2}{4N}M(N,J=N/2),
\label{eq:topA}
\ee
and
\bee
\frac{3B}{N} = M(N,J=3/2)-M(N,J=1/2) 
\label{eq:diff31}
\ee
or
\bee
B= M(N,J=N/2)-M(N,J=N/2-1).
\label{eq:topdiff}
\ee
But, these relations basically amount to fits of two states' masses to two parameters ($m_0$ and $B$).
Such a fit has  no degrees of freedom,
so there is nothing like a chi-squared parameter to tell whether the fit was good or not.
With three flavors and for
$N=3$, in the limit of degenerate $u$ and $d$ quark masses,
 there are eight relevant baryons (four members
of the octet and four members of the decuplet), and  the $1/N^2$ large-$N$ mass formula
(which includes all three-body operators) has
8 parameters, so there are unique combinations of masses which pick out particular coefficients in the mass formula.
However, again one cannot assign any quality measure to the determination of a parameter.
 And for $N>3$, the number
of baryons, even when only a selection of masses are computed,
 increases much faster than (presumably) the number of parameters. So, I will usually look at fits in which there
are more masses than parameters, and show coefficients from these fits, rather than selecting particular states to
give particular parameters in a mass formula.

The outline of the paper is as follows: In Sec.~\ref{sec:details} I briefly recapitulate how the lattice simulations
are performed, and show some representative pictures of spectroscopy.
Then, in Sec.~\ref{sec:models} I review expectations from a quark model and from large-$N$ counting.
Sec.~\ref{sec:results} presents results in the context of these models. Some conclusions are given in Sec.~\ref{sec:conclusions}.

\section{Methodology and representative results (without analysis)\label{sec:details}}

The simulations are entirely straightforward, and as nearly all the details are identical to what was presented
in Ref.~\cite{DeGrand:2012hd}, readers are referred to that work for fuller explanations.
Simulations use the usual Wilson plaquette gauge action, with clover fermions
 with normalized hypercubic (nHYP) smeared links
as their gauge connections\cite{Hasenfratz:2007rf}. The bare quark mass is introduced into the simulation
in the usual way,  via the hopping parameter $\kappa$.
The clover coefficient is fixed at its tree level value,
$c_{SW}=1$. The code is
 a version of the publicly available package of the MILC collaboration~\cite{MILC}.
As already remarked, the gauge groups are $SU(N)$ with $N=3$, 5, 7.
 The simulation volumes were all $16^3\times 32$ sites.
The bare gauge couplings were (roughly) matched so that pure gauge observables were the same on all three $N$'s;
this was done in an attempt to match discretization and finite volume effects.
The observable chosen to match was
the shorter version of the Sommer parameter \cite{Sommer:1993ce}
$r_1$, defined in terms of the force $F(r)$ between static quarks,
$r^2 F(r)= -1.0$ at $r=r_1$. The real-world value is $r_1= 0.31$ fm \cite{Bazavov:2009bb},
and with it the common lattice spacing is about 0.08 fm.
Simulation parameters are reported in Table \ref{tab:par}. 

\begin{table}
\begin{tabular}{c c c c}
\hline
                 & $SU(3)$   & $SU(5)$   &   $SU(7)$ \\
\hline
$\beta$          & 6.0175  & 17.5 & 34.9 \\
configurations   &  80     & 120   & 160   \\
$r_1/a$          & 3.90(3) & 3.77(3) & 3.91(2) \\
 \hline
 \end{tabular}
\caption{Parameters of the simulations.
\label{tab:par}}
\end{table}

Tables of the masses of states with two degenerate flavors are presented elsewhere,
in Refs.~~\cite{DeGrand:2012hd} and \cite{Cordon:2013}. The latter work includes additional states from the former:
lighter mass baryons for all $N$'s, more $N=3$ states to be used in matching data
 from different $N$'s,
and the $J=1/2$ baryons for all but the lightest two
 quark masses for $N=7$.

\begin{figure}
\begin{center}
\includegraphics[width=0.7\textwidth,clip]{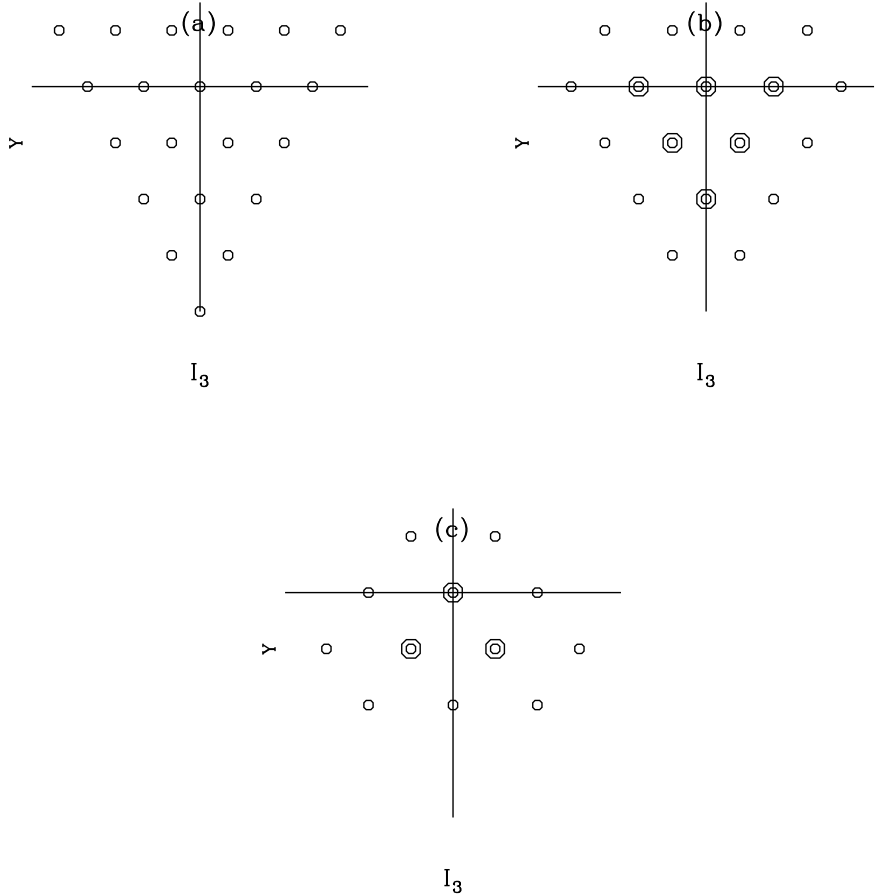}
\end{center}
\caption{Weight diagram for five color states, for (a) $J=5/2$, (b) $J=3/2$,
(c) $J=1/2$.
The horizontal axis in all cases is $Y = +2/3$.
\label{fig:weight5}}
\end{figure}

\begin{figure}
\begin{center}
\includegraphics[width=0.7\textwidth,clip]{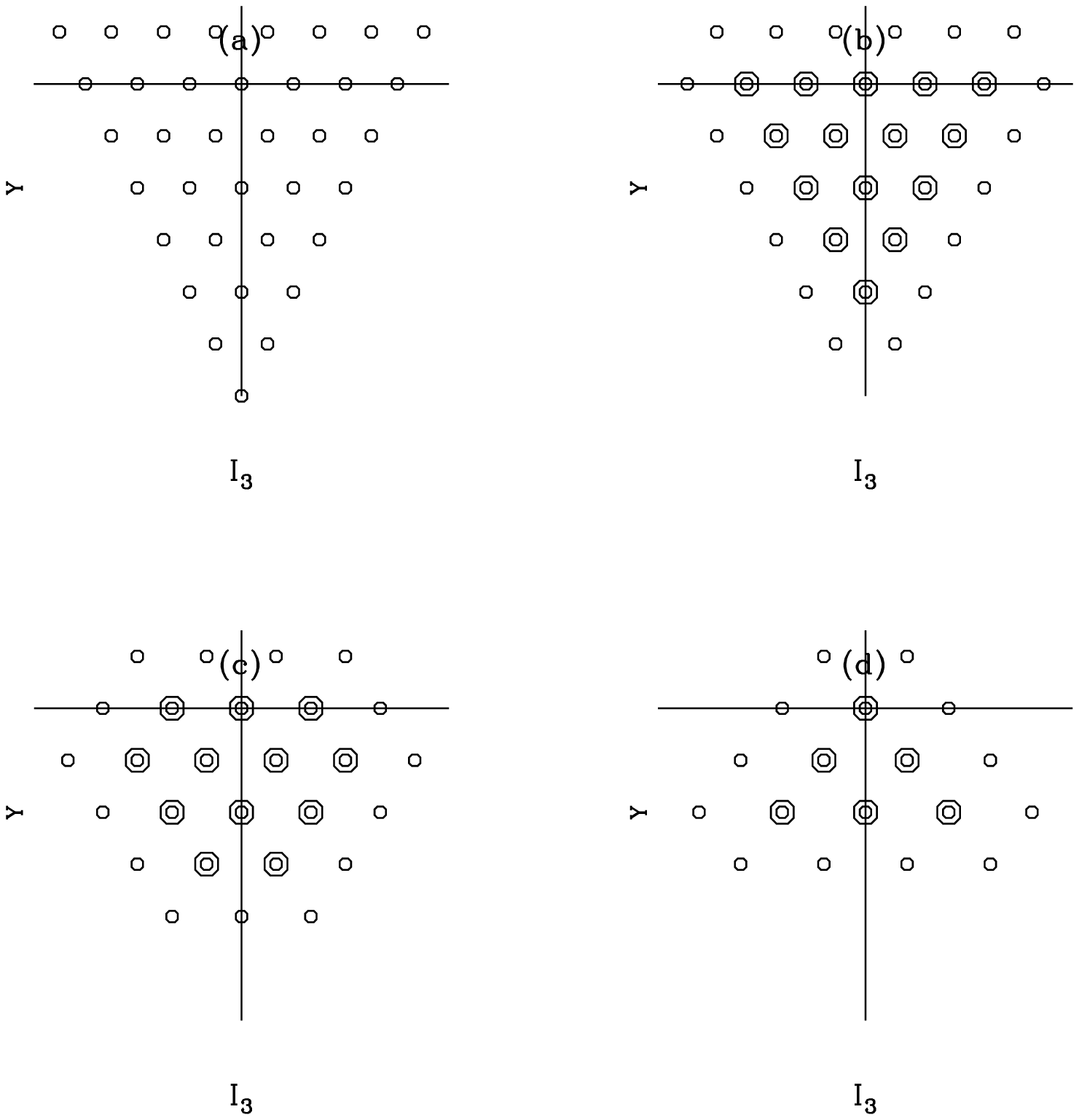}
\end{center}
\caption{Weight diagram for seven color states, for (a) $J=7/2$, (b) $J=5/2$,
(c) $J=3/2$ and (d) $J=1/2$.
The horizontal axis in all cases is $Y = +4/3$.
\label{fig:weight7}}
\end{figure}

The interpolating fields for baryons use operators which are diagonal
 in a $\gamma_0$ basis -- essentially, they create
nonrelativistic quark model trial states.
They are constructed straightforwardly in three steps. It begins with the weight
diagrams for the multiplets, displayed for $N=5$ and 7 in Figs.~\ref{fig:weight5} and \ref{fig:weight7}.
The axes are isospin $I$ and hypercharge (baryon number plus strangeness). Double zeroes show multiple states.
There are many states! But, as in the case of the ordinary $N=3$ baryons, only a few of them are interesting. In the
degenerate $u$ and $d$ mass limit, all the states with the same total isospin are mass-degenerate.
So only one state per row need be computed -- the states along the edges of the diagram are the easiest
to write down. The same consideration 
applies to the ``interior'' states (analogs of the $\Lambda$ hyperon). In the limit of exact isospin
invariance, all the members of an interior row are mass-degenerate, too.

States on the exterior of the weight diagram are easily constructed. 
We imagine constructing states which are space-symmetric. Then the
spin wave function for $n$ identical quarks has total angular momentum $J=n/2$ by symmetry and the
wave function for a hadron made of two flavors is quickly constructed, by appropriately adding the angular momenta.
Three-flavor states can be constructed by raising or lowering in $U-$ or $V-$spin.
 Wave functions for the interior states  can also be written down by inspection:
the total (and highest third component of) isospin of an ``interior-edge'' state gives the number of light
quarks coupled to nonzero angular momentum, and the extra light quarks are a set of $ud$ pairs coupled
to $(I=0$, $J=0$).

Next, we want a standard form for the interpolating field, to pass to the computer code.
 Trading the minus signs associated with the Grassmann
nature of the fermion creation operator with  color anti-symmetrization,  terms in the operator we just wrote
down can be combined, so that a
 generic three-flavor baryon interpolation field can be written as
\bee
O_{B}= \epsilon^{a_1 a_2 a_3 \dots a_N} \sum_{\{s_j\}} C_{\{s_j\}} u^{a_1}_{s_1} u^{a_2}_{s_2} \dots d^{a_k}_{s_k}\dots
d^{a_l}_{s_l}\dots s^{a_N}_{s_N}
\label{eq:bar}
\ee
where the $C$'s are an appropriate set of coefficients.

Finally, the baryon correlator must include all nonzero contractions of creation operators at the source and annihilation operators
at the sink. For each flavor, 
this gives a determinant of quark propagators, so that a term in a three-flavor baryon correlator is a
product of three sub-determinants (one for each flavor). This must be summed over all the ways that colors
can be apportioned between the quarks.

For the top of the multiplet, this is a small number of terms. For example, it is a single
$N\times N$ determinant for the corners of the $J=N/2$ multiplet. But the number of terms increases
dramatically as one moves down in $J$. For example, the $N=7$ baryon with  $J=I=1/2$  has about 1.5 million
determinant products. This is the source of the remark in the Introduction, that lower $J$ is more expensive
than higher $J$. An obvious way to lower the cost of the lower $J$ states is to keep fewer color combinations
than the full sum over all possibilities. I have played with that a bit, but all of the truncations I did
resulted
in signals which were too noisy to be useful: the error on the mass differences became large.
 This is an obvious loose end in the project.

So, I computed all eight states of the $SU(3)$ octet and decuplet (the $\Delta$, $\Sigma^*$, $\Xi^*$,
$\Omega$, p, $\Sigma$, $\Lambda$, $\Xi$). I computed all 20 states of $SU(5)$, the rightmost
 inner and outer states in Fig.~\ref{fig:weight5}. I constructed all 38 of the ``edge states
of $SU(7)$, just to count their cost. In the end I only collected the masses of the $J=7/2$ multiplet,
the ``outer'' part of the $J=5/2$ multiplet, the strangeness -1 and -5 $J=5/2$ ``inner'' states,
 and the nonstrange $J=3/2$ and $1/2$ states, 19 states in all.


\begin{figure}
\begin{center}
\includegraphics[width=0.8\textwidth,clip]{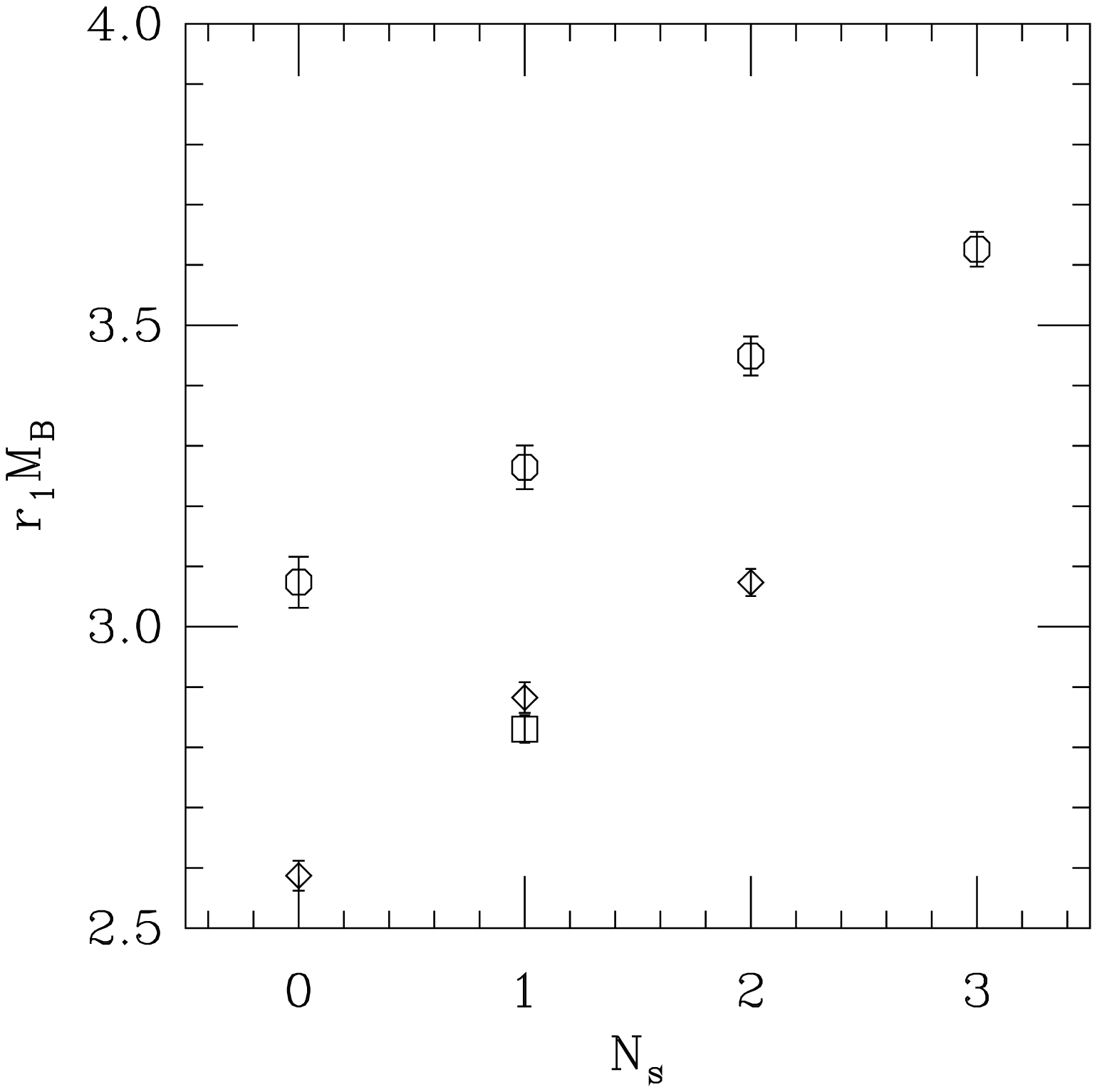}
\end{center}
\caption{A typical $SU(3)$ data set ($\kappa_l=0.1257$, $\kappa_h=0.124$)
showing the four $J=3/2$ states as octagons, the  $J=1/2$ octet as diamonds, and the $\Lambda$ as a square.
\label{fig:su3ex}}
\end{figure}

\begin{figure}
\begin{center}
\includegraphics[width=0.8\textwidth,clip]{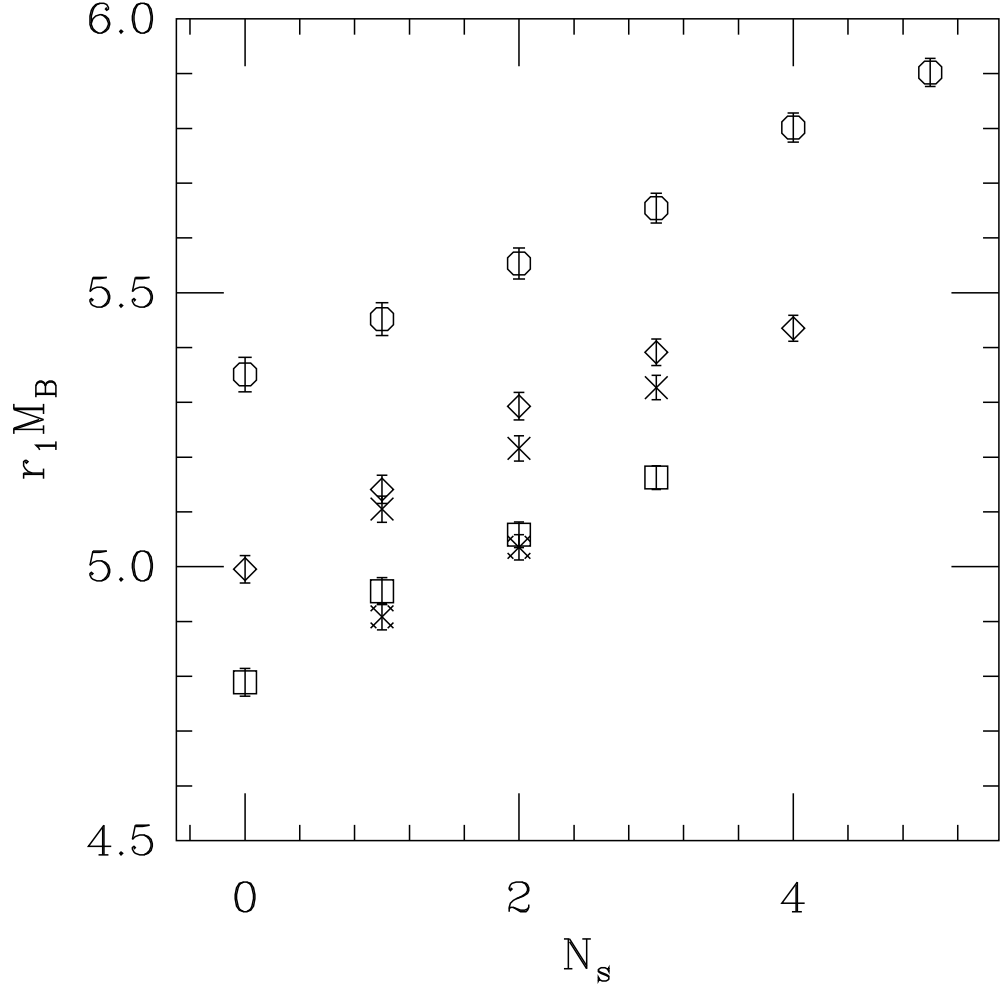}
\end{center}
\caption{A typical $SU(5)$ data set ($\kappa_l=0.127$, $\kappa_h=0.126$)
showing the  $J=5/2$ states as octagons, the  $J=3/2$ outer states as diamonds, the
$J=3/2$ inner states as crosses, the $J=1/2$ outer states as squares, and the $J=1/2$ inner 
states as fancy crosses.
\label{fig:su5ex}}
\end{figure}

\begin{figure}
\begin{center}
\includegraphics[width=0.8\textwidth,clip]{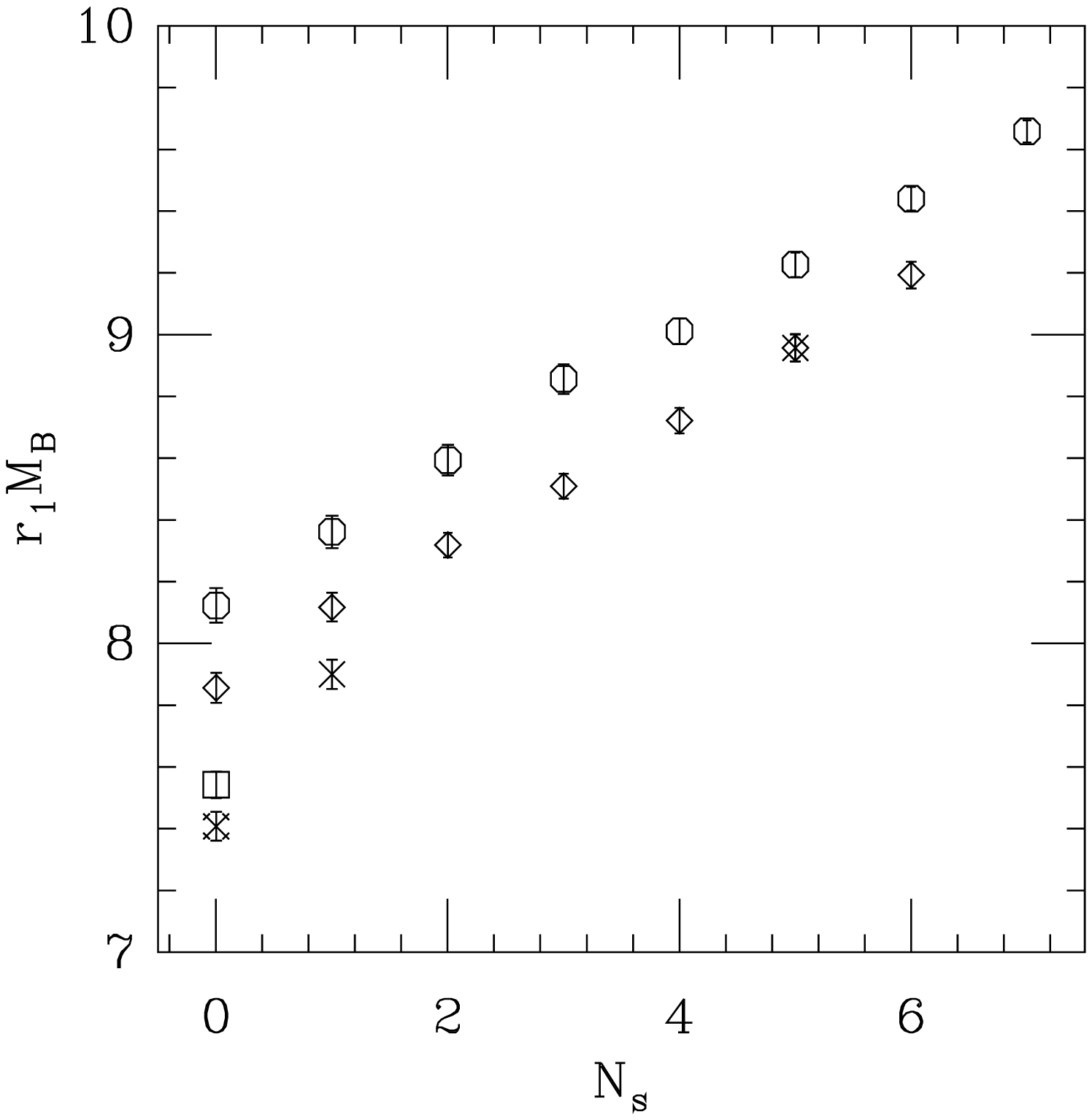}
\end{center}
\caption{The states of a  typical $SU(7)$ data set ($\kappa_l=0.129$, $\kappa_h=0.127$)
showing the $J=7/2$ states as octagons, the outer $J=5/2$ states as diamonds, 
the $J=5/2$ inner states as crosses, the $J=3/2$ state as a square, and the $J=1/2$ state as a fancy cross.
(Recall that I did not compute the masses of all the $N=7$ baryons.)
\label{fig:su7ex}}
\end{figure}

Typical results are shown in Figs.~\ref{fig:su3ex}, \ref{fig:su5ex} and \ref{fig:su7ex}.
They are remarkably similar. In all cases, there is an obvious linear rise in mass with the number of strange
quarks. In all cases, the states are ordered in mass with their angular momentum, with higher $J$ lying higher.
And in all cases, an interior state with the same strange quark content and same $J$
 as an exterior state lies slightly
lower in energy, just as the $\Lambda$ is slightly lighter than the $\Sigma$.

Further analysis requires a pause to discuss models.

\section{Models and expectations\label{sec:models}}
\subsection{The color hyperfine interaction model}
We begin with a model -- the ``color hyperfine interaction''
 picture of De Rujula, Georgi and Glashow \cite{De Rujula:1975ge},
also implemented in the bag model \cite{DeGrand:1975cf}.
The Hamiltonian for a bound state of $N_i$ quarks of constituent mass $m_i$ and spin $\vec S_i$,
all in identical S-wave spatial wave functions, is posited to be
\bee
H = \sum_i N_i m_i + \sum_{i\ne j} C_{ij}\vec S_i\cdot \vec S_j.
\ee
Here $C_{ij}$ is a mass-dependent constant representing a magnetic hyperfine interaction between the quarks.
Specializing to the case of degenerate $u$ and $d$ quarks with
constituent masses $m_u=m_d=m_0$ and a heavier strange quark of
$m_s=m_0 + \delta m$, there will be three distinct  hyperfine  coefficients, $C_{nn}$ between two nonstrange
quarks, $C_{ns}$ involving a strange and a nonstrange quark, and $C_{ss}$ for two strange quarks.
Completing the square by summing the spins for the strange and nonstrange quarks $\vec J_q= \sum_{i\in q}\vec
S_i$, the mass of a hadron will be
\bee
M = N m_0 + N_s \delta m + \frac{C_{nn}}{2}[J_n(J_n+1)-\frac{3}{4}N_n] +  C_{ns}\vec J_n\cdot \vec J_s +
\frac{C_{ss}}{2}[J_s(J_s+1) -
\frac{3}{4}N_s].
\ee
Now, because $C_{ij}$ is proportional to the product of the magnetic moments $\mu_i$ and $\mu_j$ of quarks of type
$i$ and $j$, and because $\mu_i \sim g/m_i$, $C_{ij}$
 scales like $g^2$ or like $1/N$ since $g^2$ is $1/N$ times the 't Hooft
coupling $\lambda$. 
Finally, the hyperfine interaction interpolates in the number of strange quarks, so
if we write $C_{nn}=B/N$, then $C_{ns}=(B-\delta)/N$ and $C_{ss}=(B-2\delta)/N$. Our model mass formula is
\bee
M = Nm_0 + C +  N_s \delta m + \frac{B}{N}J(J+1) + \frac{\delta}{N}[J_n(J_n+1)-J(J+1)-J_s(J_s+1)]
\ee
We can groom this if we realize that the wave function of the individual quarks is spin-isospin
or spin-flavor locked, so $J_s=N_s/2$ and $J_n=I$ the isospin. Then the mass of a baryon
with N colors, total spin $J$, isospin $I$ and containing $N_s$ strange quarks is
\bee
M(N,J,I,N_s) = Nm_0 + C +  N_s \delta m + \frac{B}{N}J(J+1) + \frac{\delta}{N}[I(I+1)-J(J+1) -
\frac{N_s}{2}(\frac{N_s}{2} +1)].
\label{eq:colorspin}
\ee
The four constants $m_0$, $\delta m$, $B$ and $\delta$ are presumably functions of the nonstrange quark mass
and $\delta m$ and $\delta$ are additionally functions of the strange quark mass. As we will see, this model
encodes many of the regularities in lattice spectral data.

\subsection{Large-$N$ parametrizations}
A model is just a model, but in the context of $1/N$ expansions one can do better -- one can
write down general expressions
for the mass Hamiltonian as a sum of coefficients of powers of $N$
\cite{Jenkins:1995td,Dai:1995zg}. Each coefficient represents
the expectation value of an $n$-quark interaction. The dependence on $J$, $I$, and $N_s$ is fixed by symmetry.
There are two relevant situations. The simplest one to use assumes only isospin symmetry and is labeled 
an $SU(2)\times U(1)$ mass formula in the literature:
\beea M &=& Nm_0 + C +  N_s \delta m + \frac{1}{N}[c_1 J(J+1)+c_2 I(I+1)+c_3(\frac{N_s}{2})^2] \nonumber \\
& &+ \frac{1}{N^2}\frac{N_s}{2}[c_4 J(J+1)N_s + c_5 I(I+1)(\frac{N_s}{2})^2 + c_6(\frac{N_s}{2})^2]
+\dots \nonumber \\
\label{eq:su2u1}
\eea
All the coefficients are (in principle arbitrary) functions of the nonstrange and strange quark masses.

There is also a formula derived assuming that flavor $SU(3)$ is softly broken. With some abuse of notation, and
working to
order $1/N$, it is
\beea M &=& Nm_0 + C + \delta m (N-3N_s) +  \frac{1}{N}[d_1 J(J+1) \nonumber \\
& & + d_2[3I(I+1) - J(J+1) -\frac{3N_s}{2}(\frac{N_s}{2}+1) ] + d_3 \frac{(N-3N_s)^2}{6} ]
+\dots \nonumber \\
\label{eq:su3}
\eea
The terms $\delta m$ and $d_2$ are linear in the flavor $SU(3)$ breaking parameter
 (some quantity analogous to $\delta m$ in the other mass formulas),
 while $d_3$ is proportional to $(\delta m)^2$.

In principle, the coefficients could themselves contain non-leading corrections parametrized by powers
of $1/N$. The constant term $C$ is an example of such a correction; it is not usually written down
in the continuum literature.

At order $1/N$, the color hyperfine  formula is a special case of Eq.~\ref{eq:su2u1}, with $c_1=B-\delta$,
$c_2= -c_3=\delta$. The parameters in Eqs.~\ref{eq:su2u1} and \ref{eq:su3} are also linearly related.

In the limiting case of two flavors of degenerate mass, all
 of these formulas reduce to the simple expression
\bee
 M(N,J) = Nm_0 + C + \frac{J(J+1)}{N}B .
\label{eq:jsplit}
\ee

All of these formulas share common features, which we can search for in the data.

\section{Comparison of results and expectations\label{sec:results}}
\subsection{Matching different $N$'s}
The coefficients in the mass formulas, Eqs.~\ref{eq:colorspin}, \ref{eq:su2u1}, and \ref{eq:su3}, 
are implicitly functions
of the masses of the nonstrange and strange quarks. In a lattice simulation, these masses are ultimately
the (lattice-regulated) bare quark masses which are inputs to the simulation. These bare masses,
 like the bare gauge couplings,
are different for the different $N$'s which are simulated. So, how are we to compare data at different $N$?

This is a question which does not have a unique answer, but a reasonable approach is to try to match the values of some 
dimensionless, non-baryonic observable across $N$.
Since the gauge sector was already used to match lattice spacings, and since I will continue to use a gauge 
observable, $r_1$, to set the lattice spacing, ``matching'' is a shorthand for ``find the value of $\kappa$'s from all
three $N$'s at which the chosen observable is nearly equal.''

 This can be done in two stages: First, one can try to match 
the nonstrange quark 
mass, which can be done using the spectroscopy of states which are built of degenerate up and down quarks.
Then I will not try to match the strange quark mass, but present plots showing fit parameters as a function of
$\delta m$.

\begin{figure}
\begin{center}
\includegraphics[width=0.7\textwidth,clip]{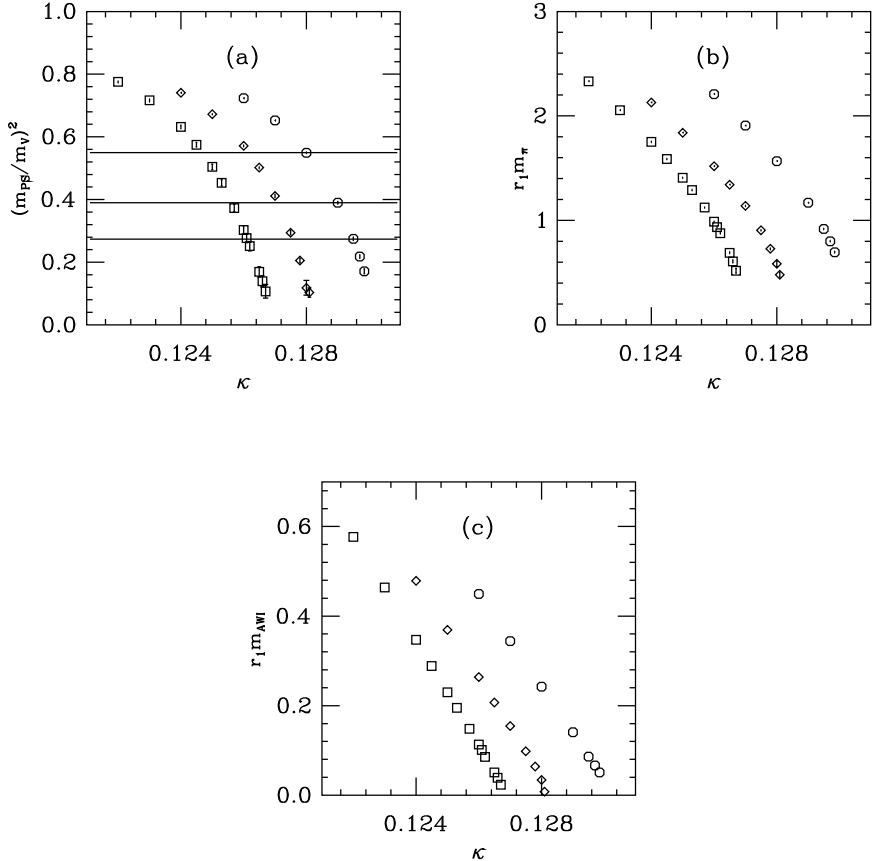}
\end{center}
\caption{Observables which might be used to match data from different $N$'s: (a) the
squared pseudoscalar to vector meson mass ratio (b) $r_1 m_\pi$ (c) $r_1 m_q$ where $m_q$ is the AWI quark mass.
Squares, diamonds and octagons label $N=3$, 5, and 7. In all cases the horizontal axis is the bare hopping
parameter $\kappa$. The horizontal lines in panel (a) connect $\kappa$'s which are matched for analysis.
\label{fig:match}}
\end{figure}

There are, of course, many possible quantities which can be used to do the matching. 
All will give slightly different choices for the match, simply because the theories are different.
But, let us see what the data says: In Fig.~\ref{fig:match}, I show the squared pseudoscalar to
 vector meson mass ratio, $r_1 m_{PS}$ and $r_1 m_{AWI}$ (the Axial Ward Identity quark mass) as
 a function of $\kappa$. The same
$\kappa$'s are used in all three panels. A moment with a straightedge reveals that all three observables
match reasonably well, with the same choices of $\kappa$'s.

This figure shows more data than I had in Ref.~\cite{DeGrand:2012hd}; it includes lighter quark masses to
 try to get into the chiral regime, plus additional $SU(3)$ points, for matching. The three horizontal lines
in panel (a) show the bare couplings where the nonstrange quarks will be set to study broken flavor $SU(3)$.
(They correspond to $\kappa$'s for $N=3$, 5, 7 of (0.1261, 0.1275, 0.1295), (0.1257, 0.127, 0.129),
and  (0.1261, 0.1275, 0.1295) for which $(m_{PS}/m_V)^2$ ratios are  0.27, 0.39, and 0.55.)
 Because of the inherent ambiguity in the matching procedure, I have not
tried to interpolate data to sit precisely on a matching line. I did construct additional
$N=3$ spectroscopy to fill in gaps on the matching line.

\subsection{Revisiting flavor $SU(2)$}
\begin{figure}
\begin{center}
\includegraphics[width=\textwidth,clip]{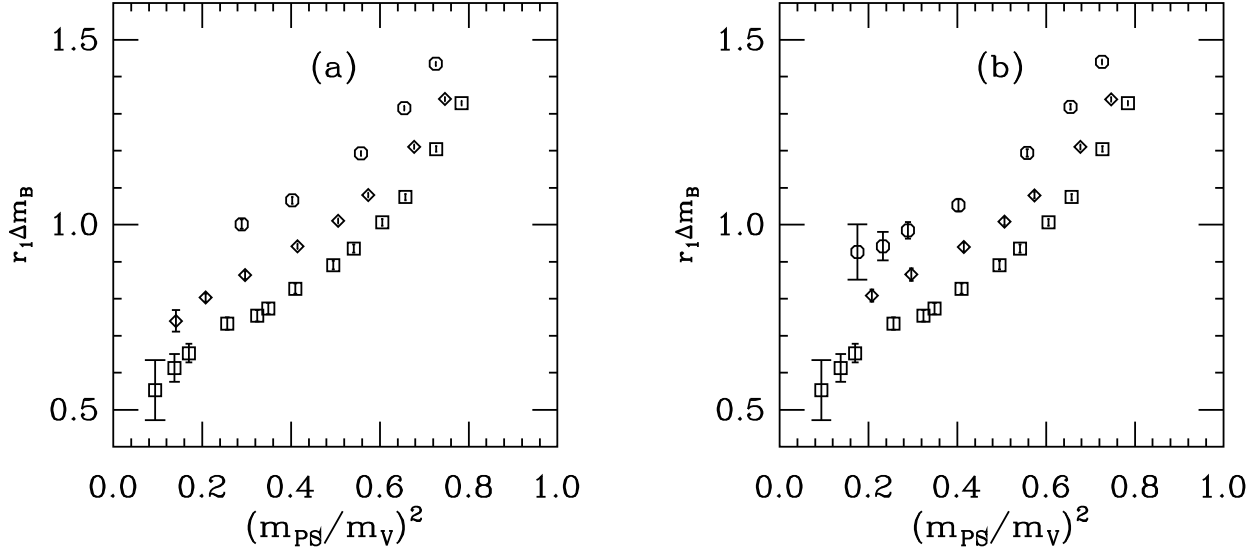}
\end{center}
\caption{
The parameter $m_0$ from  two-flavor degenerate mass data as a  function of $(m_{PS}/m_V)^2$:
(a) Data from Eq.~\protect{\ref{eq:botA}}.
(b) The $J=N/2$ vs $J=N/2-1$ mass difference of Eq.~\protect{\ref{eq:topA}}.
Data from the $SU(3)$, $SU(5)$, and $SU(7)$ multiplets are
shown respectively
as squares, diamonds, and octagons.
\label{fig:Aterm}}
\end{figure}

Before introducing the strange quark, it is necessary to return to the flavor $SU(2)$ case. Most of
this revisit involves correcting a plotting error in Ref.~\cite{DeGrand:2012hd}, where the analysis left out
the $C$ term of Eq.~\ref{eq:jsplit}. Fig.~\ref{fig:Aterm} is a corrected plot, overlaying the mass differences
of Eqs.~\ref{eq:botA} and \ref{eq:topA}, which reveal that $m_0$ drifts with $N$ at fixed  pseudoscalar - vector
mass ratio values. 

\begin{figure}
\begin{center}
\includegraphics[width=0.7\textwidth,clip]{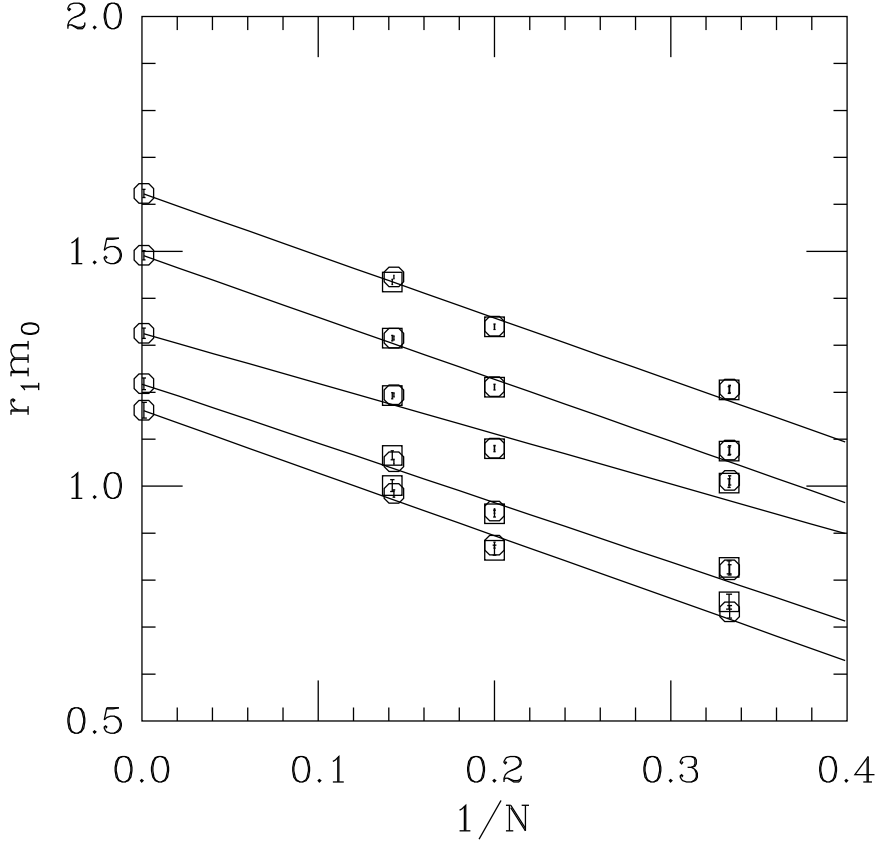}
\end{center}
\caption{The $m_0$ parameter from ``matched'' data sets, scaled by $r_1$, as a function of $1/N$.
Although they cannot be distinguished, the octagons are $m_0$ values from a fit to 
Eq.~\protect{\ref{eq:jsplit}} 
while the squares are from the scaled mass difference Eq.~\protect{\ref{eq:diff31}}.
Lines are simple linear fits in $1/N$.
The points at $1/N=0$ show the extrapolated values of $r_1 m_0$.
\label{fig:m0vs1n}}
\end{figure}

What is happening is that fits to mass formulas at fixed $N$ cannot resolve a difference between
$N m_0$ and $N(m_0 + C/N)$ -- both terms come from the $J$-independent part of the mass.
To see that this is the (approximate) explanation for the drift, take data which is matched, compute
$m_0$, and plot it as a function of $1/N$. This is shown in Fig.~\ref{fig:m0vs1n}. Here the data comes from a fit
 to Eq.~\ref{eq:jsplit} at each individual $N$ 
of all the $J$ states at a given $\kappa$. These fits all have excellent $\chi^2/DoF$
for $N=5$ and 7, where there is a nonzero number of degrees of freedom (and $\chi^2=0$ for $N=3$).
It's clear that the data show a dominantly linear $1/N$ dependence. The slope of this line is the $C$ parameter of
Eq.~\ref{eq:jsplit}.

\begin{figure}
\begin{center}
\includegraphics[width=\textwidth,clip]{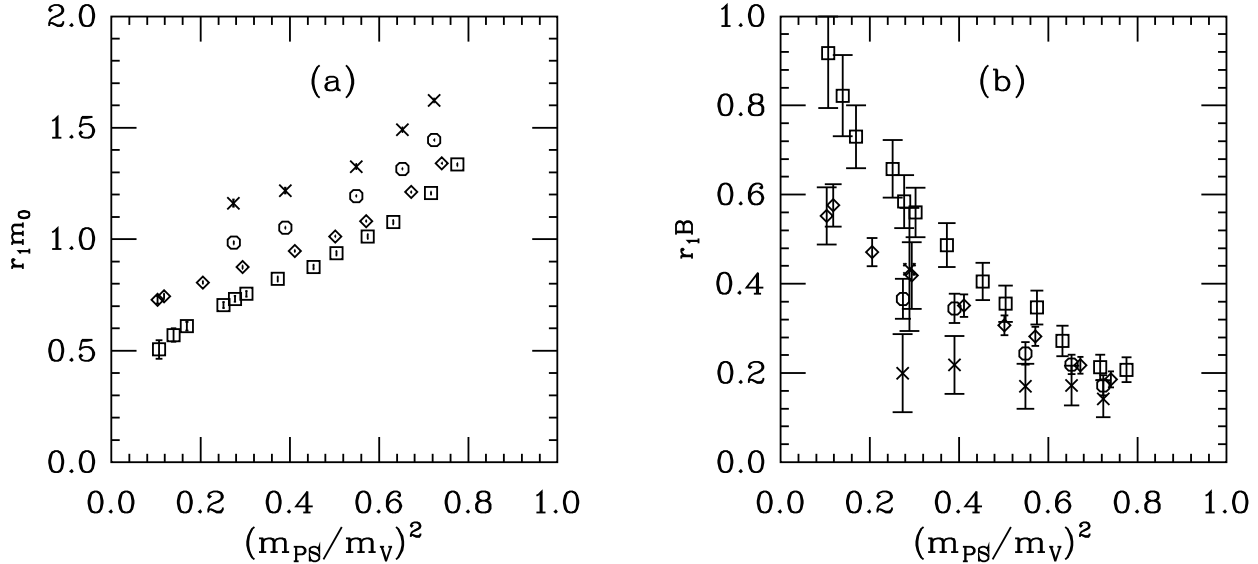}
\end{center}
\caption{
The parameter $m_0$ (panel (a)) and $B$ (panel (b)
from  two-flavor degenerate mass data as  function of $(m_{PS}/m_V)^2$ from a fit to 
Eq.~\protect{\ref{eq:jsplit}} .
Data from the $SU(3)$, $SU(5)$, and $SU(7)$ multiplets are
shown respectively
as squares, diamonds, and octagons. Crosses show extrapolations to $1/N=0$.
\label{fig:pvspirho}}
\end{figure}

\begin{figure}
\begin{center}
\includegraphics[width=0.7\textwidth,clip]{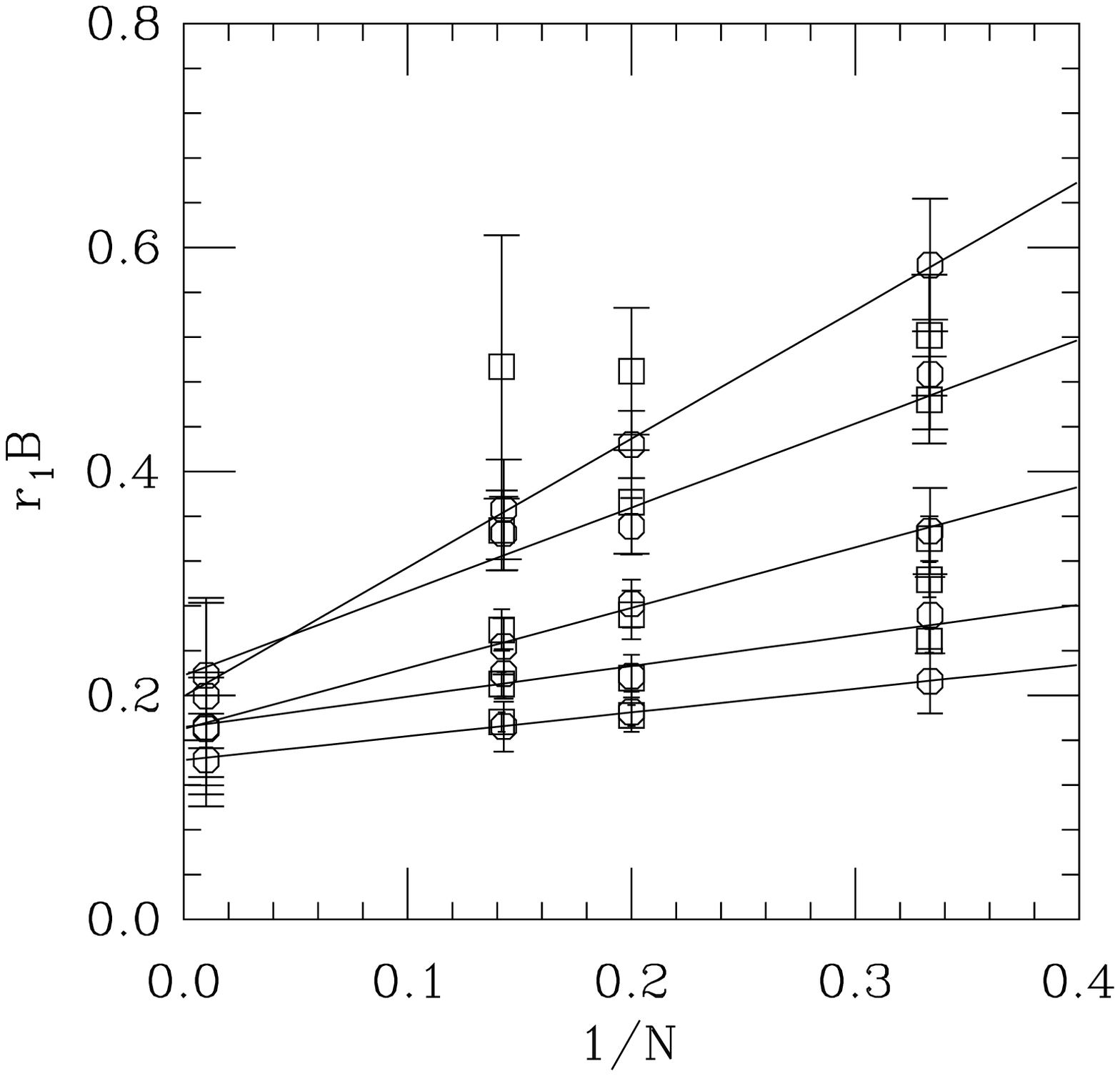}
\end{center}
\caption{The $B$ parameter from ``matched'' data sets, scaled by $r_1$, as a function of $1/N$.
The octagons are $B$ values from a fit to
Eq.~\protect{\ref{eq:jsplit}}
while the squares are from the scaled mass difference Eq.~\protect{\ref{eq:diff31}}.
The lines are fits to the octagon data of $r_1B= r_1 B_0 + r_1B_1/N$.
The points at $1/N=0$ show the values of $r_1 B_0$.
\label{fig:bvs1n}}
\end{figure}

One can repeat this analysis for the $B$ term. The individual values are shown in Fig.~\ref{fig:pvspirho}
and matched values are shown in Fig. ~\ref{fig:bvs1n}. The data are somewhat noisier than for $m_0$;
this is due to the fact that  $B$ comes from mass differences. The data reveals a non-leading in $1/N$ correction 
to $B$.

In both Figs.~\ref{fig:m0vs1n} and \ref{fig:bvs1n}, the lines  show the results 
linear fits to the matched data, $p(N)=p_0 + p_1/N$, and the points near the origin show $p_0 \pm \Delta p_0$.
 The $m_0$ fits are of poor quality: the $\chi^2$'s
range from 10 to 80. Probably this is due a $1/N^2$ or higher order term in the expansion.
 With only three
$N$'s, it is not worthwhile to look for it since a fit with an additional
 $1/N^2$ term would have no degrees of freedom.
Fits to the $B$ term have high confidence  ($\chi^2<1.0$)
 due to the larger starting uncertainties in the data. 
(This is especially the case at lighter quark masses; the largest $B$
points in Fig.~\ref{fig:bvs1n} corresponds to the lightest quark mass where I attempted
a match, and the $B$'s from Eqs.~\ref{eq:jsplit} and \ref{eq:diff31} differ  by
 a large-error-bar
1 $\sigma$.)
Fig.~\ref{fig:pvspirho} includes the extrapolations as crosses.

\subsection{Flavor $SU(3)$}
The addition of the strange quark gives the potential data set a two-dimensional nature. In order
not to overwhelm the reader with too many similar looking plots, I will proceed as follows: I will
collect data at bare parameter values where the nonstrange quark mass is matched
 (using the squared pseudoscalar to vector ratio) and present the data as a function of the parameter $\delta m$.

Again, there is some ambiguity of choice in doing a fit: for example, should one weight a state in the fit by its
$2J+1$ degeneracy, or not? I arbitrarily chose to weigh all the states equally, regardless of their
quantum numbers.

\begin{figure}
\begin{center}
\includegraphics[width=0.8\textwidth,clip]{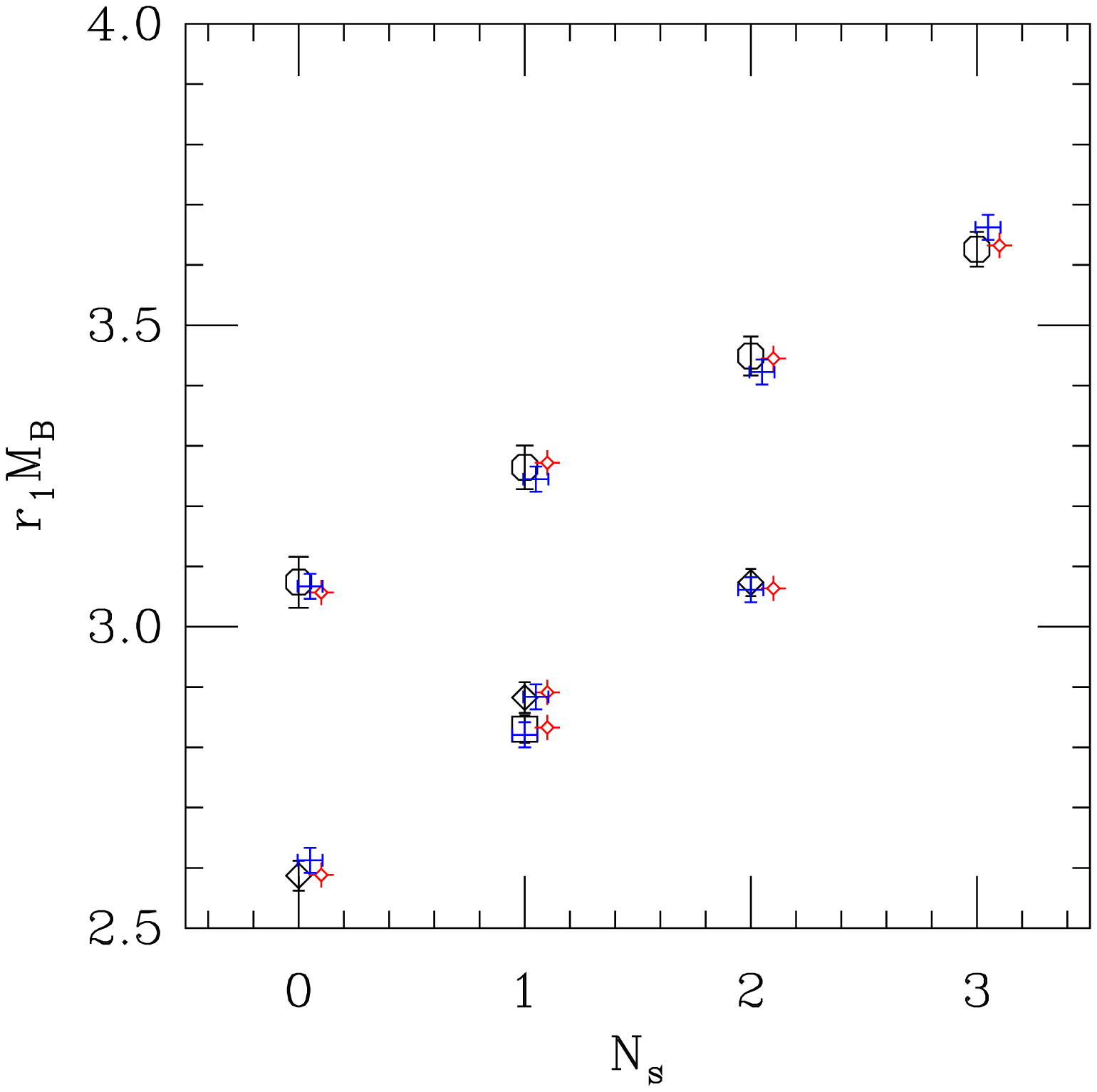}
\end{center}
\caption{Fig.~{\protect{~\ref{fig:su3ex}}} repeated, with a fit to the $SU(2)\times U(1)$ mass formula,
Eq.~{\protect{\ref{eq:su2u1}}}, overlaid as red fancy diamonds and
a fit from to the color hyperfine formula,
Eq.~{\protect{\ref{eq:colorspin}}}, overlaid as blue fancy plusses.
\label{fig:su3fit}}
\end{figure}

\begin{figure}
\begin{center}
\includegraphics[width=0.8\textwidth,clip]{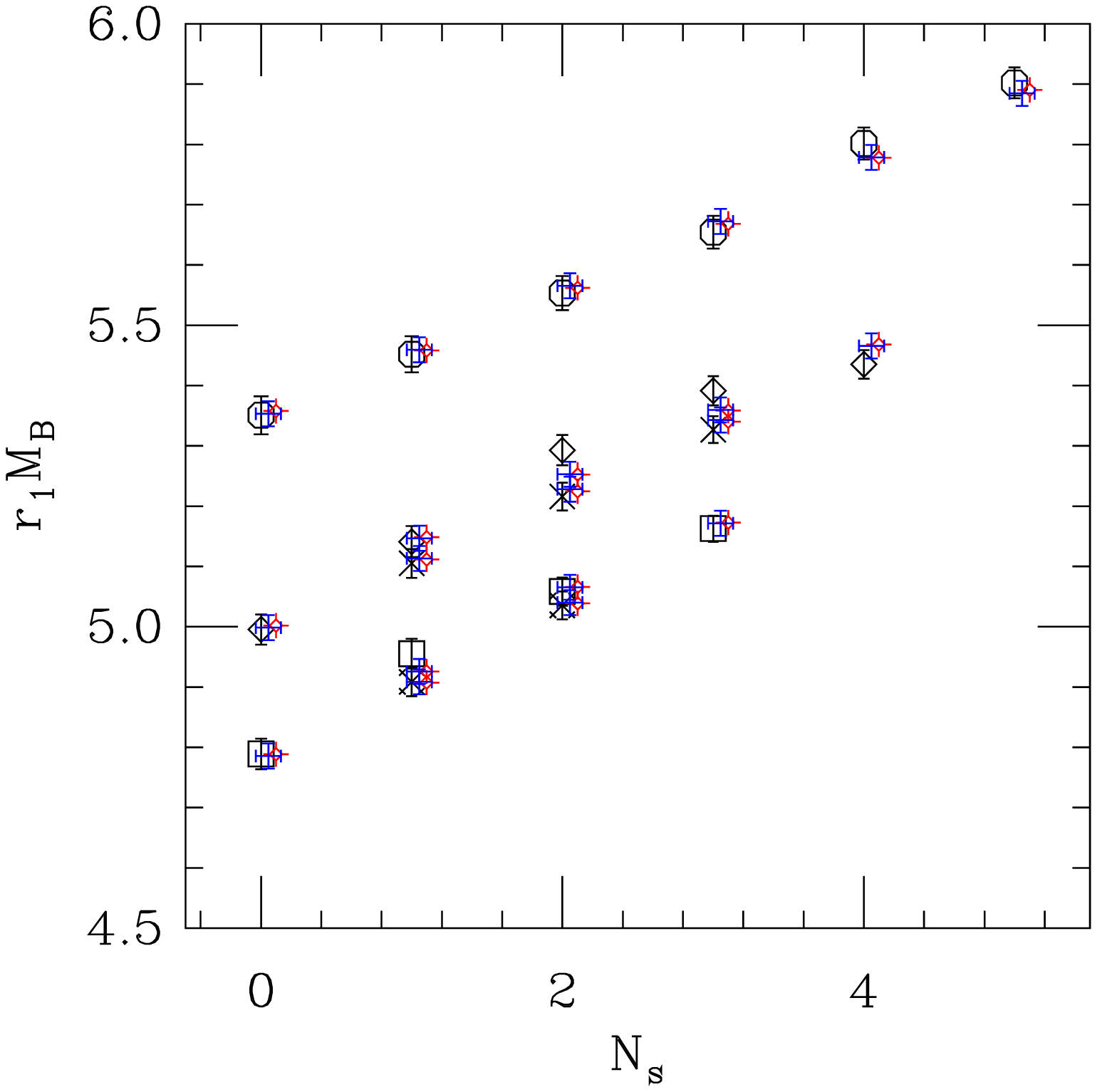}
\end{center}
\caption{Fig.~{\protect{~\ref{fig:su5ex}}} repeated, with a fit to the $SU(2)\times U(1)$ mass formula,
Eq.~{\protect{\ref{eq:su2u1}}}, overlaid as red fancy diamonds and
a fit from to the color hyperfine formula,
Eq.~{\protect{\ref{eq:colorspin}}}, overlaid as blue fancy plusses.
\label{fig:su5fit}}
\end{figure}

\begin{figure}
\begin{center}
\includegraphics[width=0.8\textwidth,clip]{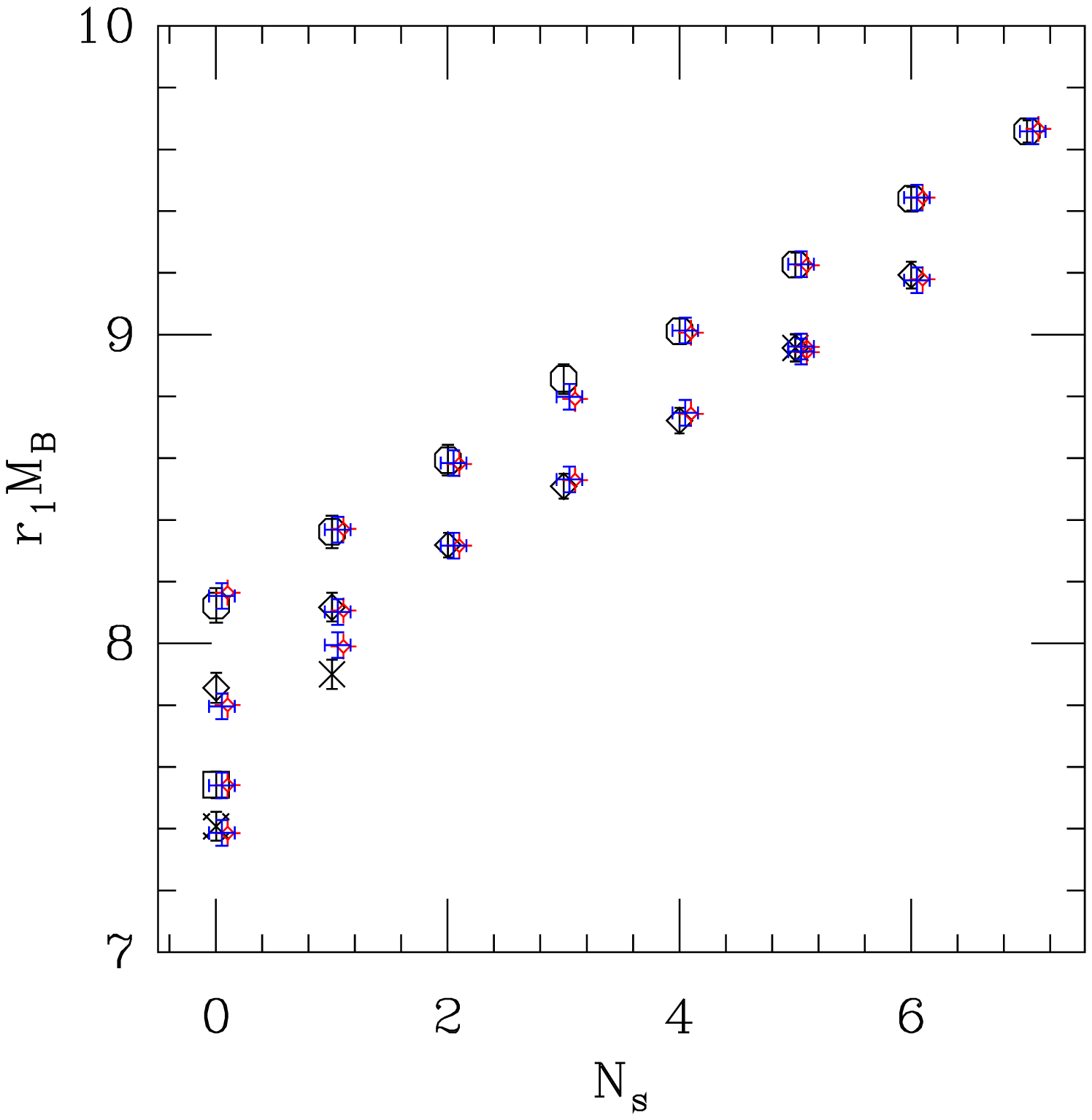}
\end{center}
\caption{Fig.~{\protect{~\ref{fig:su7ex}}} repeated, with a fit to the $SU(2)\times U(1)$ mass formula,
Eq.~{\protect{\ref{eq:su2u1}}}, overlaid as red fancy diamonds and
a fit from to the color hyperfine formula,
Eq.~{\protect{\ref{eq:su2u1}}}, overlaid as blue fancy plusses.
\label{fig:su7fit}}
\end{figure}

All fits to an individual nonstrange - strange mass combination show the following features:
\begin{itemize}
\item Fits through order $1/N$ all have $\chi^2/DoF < 1-2$. These are good fits, though the reader should recall,
 all the data come from the same underlying configurations and hence are highly correlated.
\item Fits including $1/N^2$ terms do not have appreciably lower $\chi^2$ and the fitted values of the $1/N^2$
coefficients have large uncertainties, typically much larger than their fitted values.
\item The color hyperfine mass formula is also an excellent fit to the data; it has one fewer parameter
than strict $1/N$ counting and typically only slightly higher $\chi^2$, still in the
 range below 1-2 per degree of freedom. Examples of these fits are 
shown in Figs.~\ref{fig:su3fit}-\ref{fig:su7fit}.
\end{itemize}

Therefore, I will restrict the subsequent comparisons to $1/N$ formulas. This means that the parameters
in the $SU(2)\times U(1)$
and $SU(3)$ formulas are linearly related, and the fits have the same $\chi^2$.

\begin{figure}
\begin{center}
\includegraphics[width=0.8\textwidth,clip]{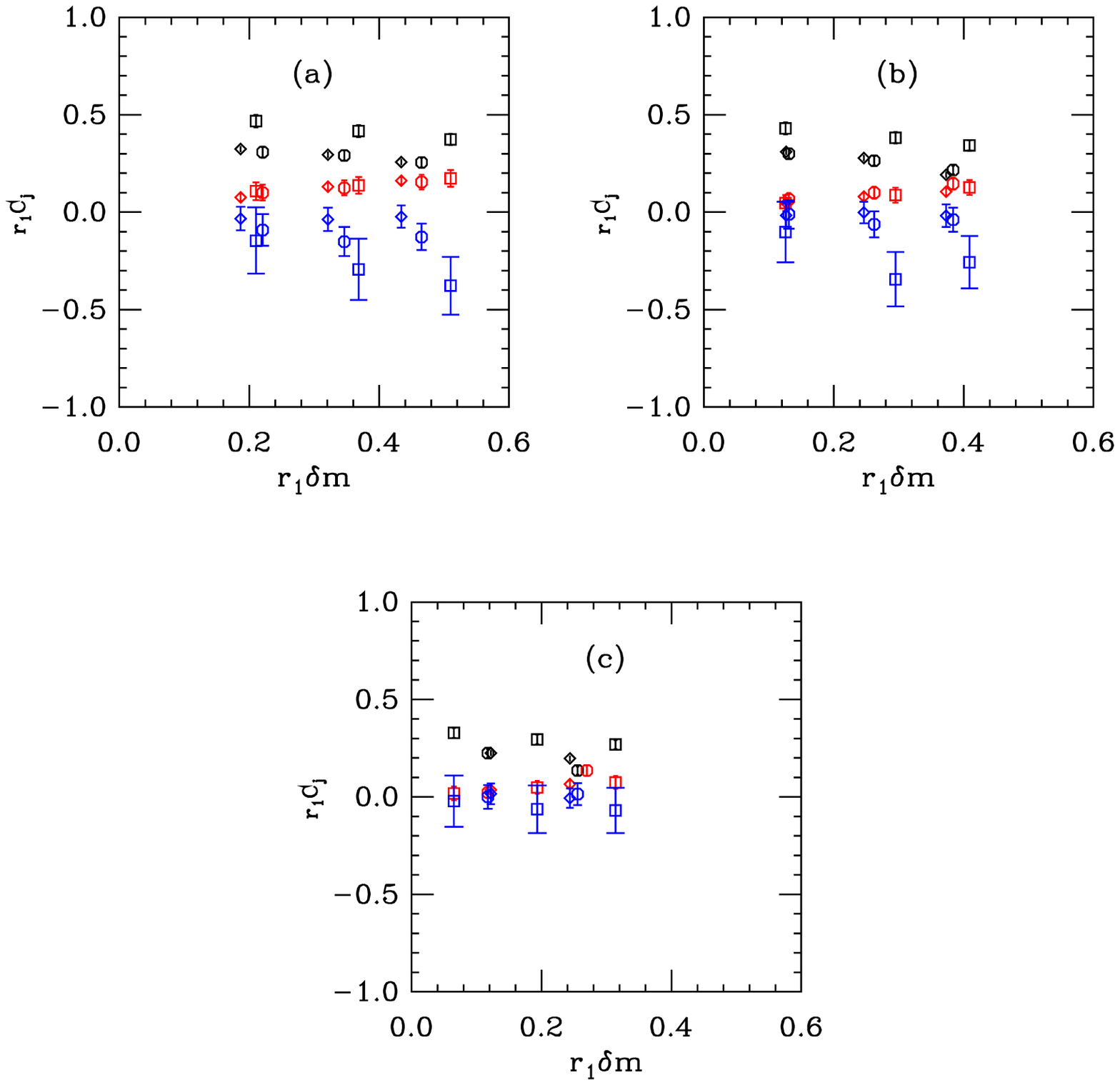}
\end{center}
\caption{
$SU(2)\times U(1)$ fits at $1/N$ to the data at matched $(m_{PS}/m_V)^2$ ratios:
(a) $(m_{PS}/m_V)^2=0.27$ (b) $(m_{PS}/m_V)^2=0.39$,  (c) $(m_{PS}/m_V)^2=0.55$.
The black points at the top of the graph are the fitted values of $c_1$ in  Eq.~{\protect{\ref{eq:su2u1}}}. The red points
in the middle
show $c_2$ and the (lowest) blue ones show $c_3$. The data are plotted as a function of $\delta m$.
Data are squares for $N=3$, diamonds for $N=5$ and octagons for $N=7$.
\label{fig:c}}
\end{figure}

\begin{figure}
\begin{center}
\includegraphics[width=0.8\textwidth,clip]{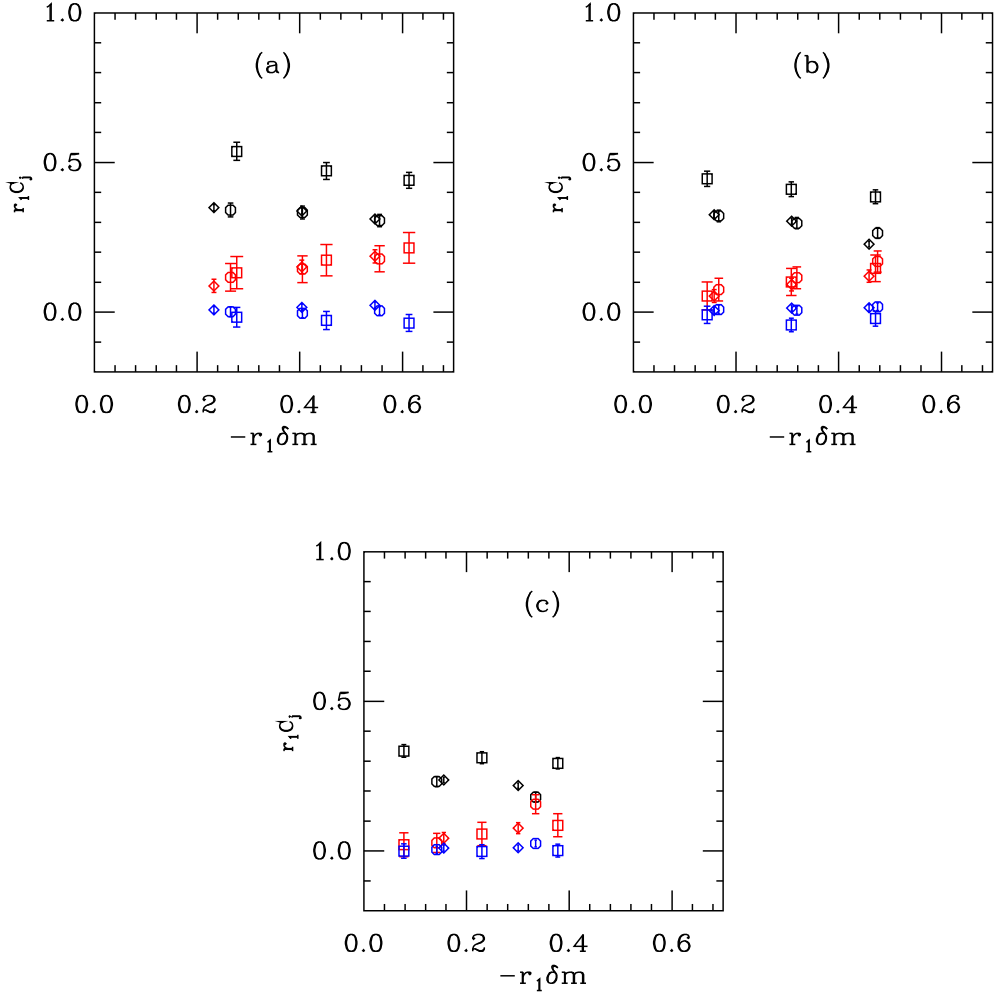}
\end{center}
\caption{$SU(3)$ fits at $1/N$ to the data at matched $(m_{PS}/m_V)^2$ ratios:
(a) $(m_{PS}/m_V)^2=0.27$ (b) $(m_{PS}/m_V)^2=0.39$  (c) $(m_{PS}/m_V)^2=0.55$.
The black points at the top of the graph are the fitted values of $d_1$ in  Eq.~{\protect{\ref{eq:su3}}}. The red points
(the middle set of points with small error bars)
show $d_2$ and the blue ones (the lowest set) show $d_3$. The data are plotted as a function of $\delta m$.
Data are squares for $N=3$, diamonds for $N=5$ and octagons for $N=7$.
\label{fig:csu3}}
\end{figure}

\begin{figure}
\begin{center}
\includegraphics[width=0.8\textwidth,clip]{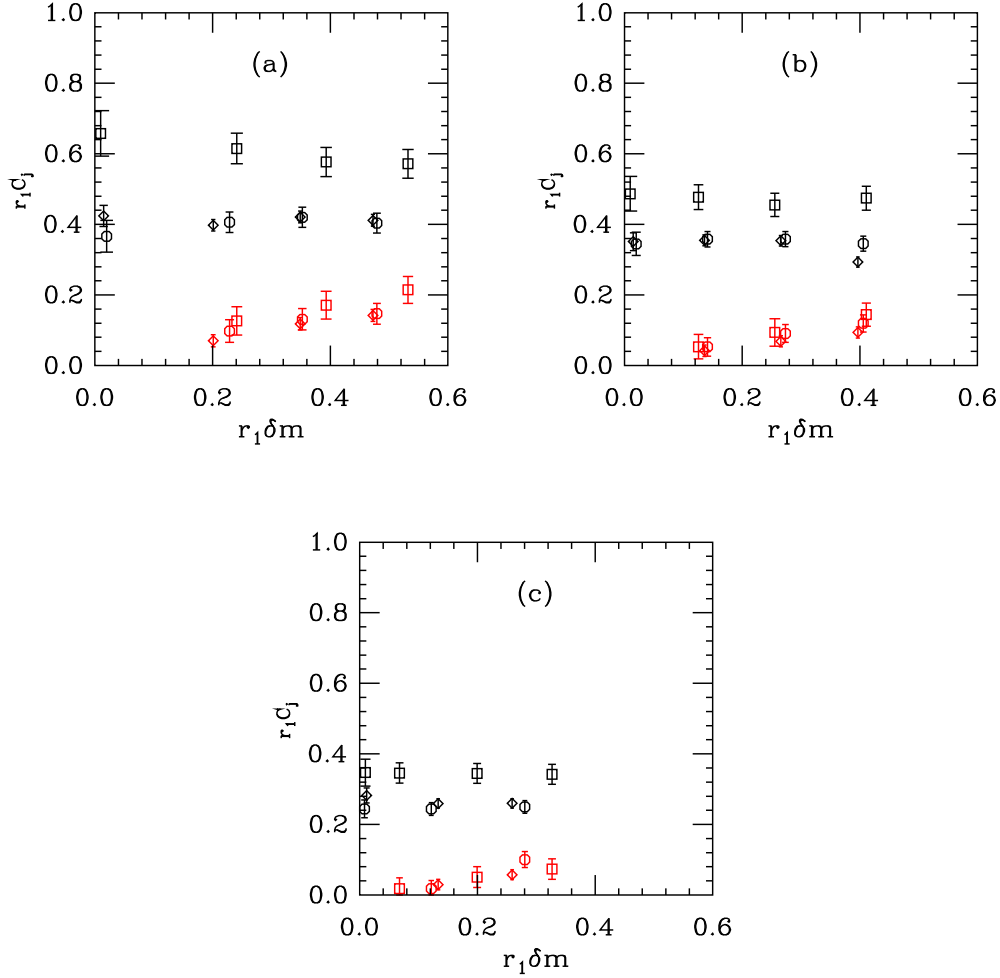}
\end{center}
\caption{Color hyperfine interaction model
 fits to the data at matched $(m_{PS}/m_V)^2$ ratios:
(a) $(m_{PS}/m_V)^2=0.27$ (b) $(m_{PS}/m_V)^2=0.39$  (c) $(m_{PS}/m_V)^2=0.55$.
The black points (the upper set) are the fitted values of $dB$ in Eq.~{\protect{\ref{eq:colorspin}}}. The red points
(the lower set) show $\delta$.  The data are plotted as a function of $\delta m$; the points at the origin show 
the values of $B$ from
a fit of the flavor $SU(2)$ data to Eq.~{\protect{\ref{eq:jsplit}}}.
Data are squares for $N=3$, diamonds for $N=5$ and octagons for $N=7$.
\label{fig:ccs}}
\end{figure}

Results of this procedure are shown for three $(m_{PS}/m_V)^2$ ratios, 0.27, 0.39, and 0.55.
Fig.~\ref{fig:c} shows the parameters from fits to the $SU(2)\times U(1)$ formula,  Eq.~\ref{eq:su2u1}.
Fig.~\ref{fig:csu3} shows the parameters from fits to  the $SU(3)$ formula, Eq.~\ref{eq:su3}.
Fig.~\ref{fig:ccs} shows the parameters from fits to the color hyperfine formula, Eq.~\ref{eq:colorspin}.

In all cases, the coefficient of $J(J+1)/N$ shows a variation with $N$ which is quite similar
 to what was seen in the two-flavor situation above. The other parameters show a striking linear dependence
on $\delta m$. Given our intuition from the color hyperfine model, this is absolutely what is expected.
The parameter $c_3$ in the $SU(2)\times U(1)$ parametrization is poorly determined by the data.

\begin{figure}
\begin{center}
\includegraphics[width=0.7\textwidth,clip]{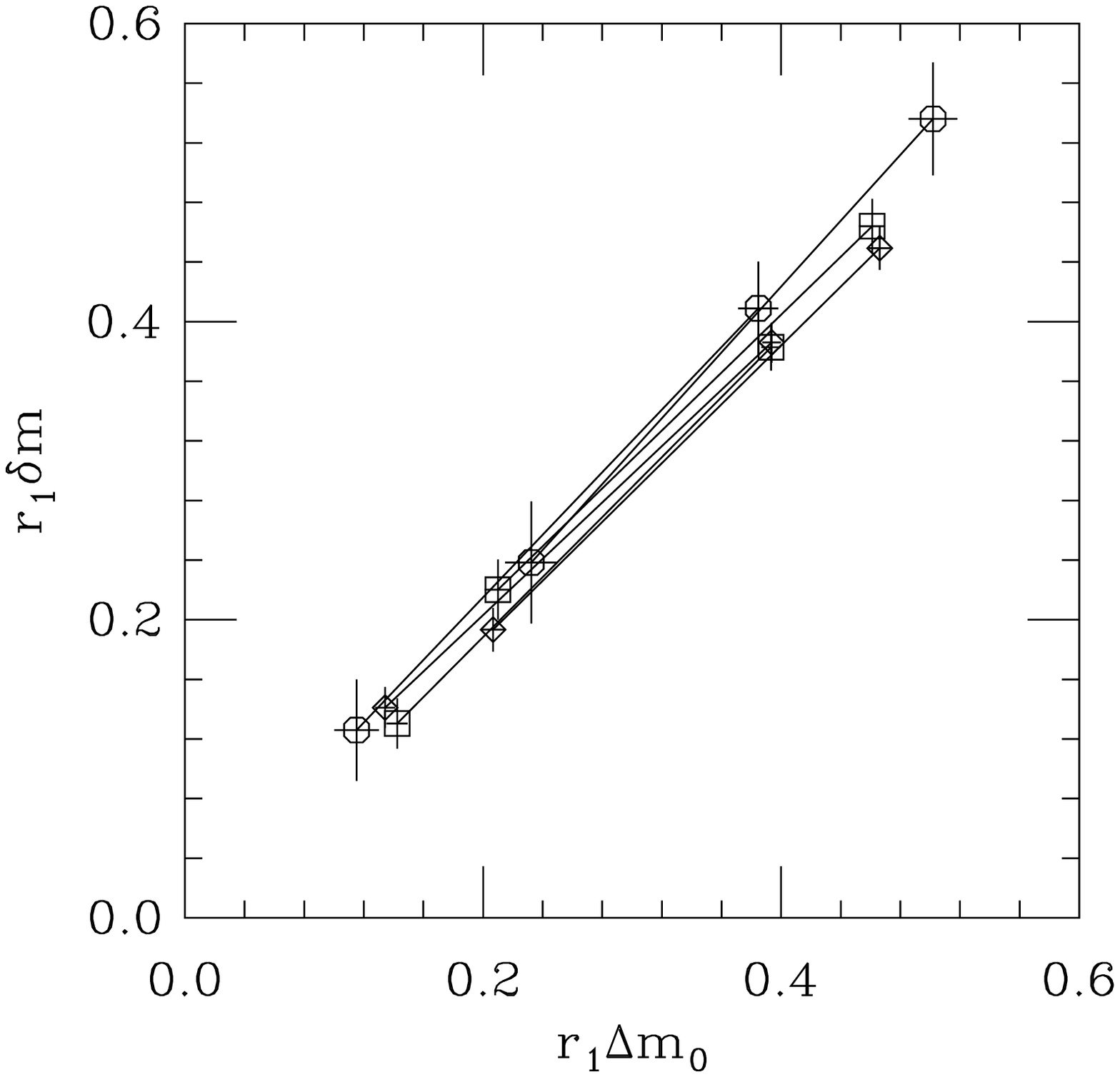}
\end{center}
\caption{Fit $\delta m$ from  Eq.~{\protect{\ref{eq:su2u1}}} as a function of the differences
in $m_0$ between the two bare masses Eq.~\protect{\ref{eq:jsplit}} at $(m_{PS}/m_V)^2=0.39$.
Data are squares for $N=2$, diamonds for $N=5$ and octagons for $N=7$.
This shows that $\delta m$ is in fact the difference in constituent quark masses.
\label{fig:dmvsdm}}
\end{figure}

In the three-flavor formulas, the parameter $\delta m$ is just a fit parameter, but of course it has
physical meaning as the difference between the constituent quark masses of the strange and nonstrange
quarks. We can test that equality by plotting $\delta m$ from fits to Eq.~\ref{eq:su2u1}
as a function of the difference of $m_0$'s
appropriate to fits to Eq.~\ref{eq:jsplit} for  the two quark masses used in the three-flavor spectroscopy.
This is shown in Fig.~\ref{fig:dmvsdm}. All is as expected.

In the color hyperfine interaction model, the parameter $B$ is supposed to be unchanged
by the presence of strange quarks. I plotted it in Fig.~\ref{fig:ccs}, near $\delta m=0$.
One might imagine that the extra states might pull $B$ away from this value, but that 
is not the case. The plot is a bit redundant, because the states with no strange quarks are
also included in the fits at $\delta m \ne 0$.

In the $SU(2)\times U(1)$ and color hyperfine formulas,
 $\delta m$ is the nonstrange quark -strange quark mass difference.
In the $SU(3)$ formula, $\delta m$ is the coefficient of $N-3N_s$ so the
 meaning of $m_0$ and $\delta m$ as constituent masses is  obscured.
 This is why the x-axis in
Fig.~\ref{fig:csu3} is $-\delta m$.

Recall that the color hyperfine Hamiltonian is a particular choice for the terms
in the $SU(2)\times U(1)$ parametrization, with $B-\delta = c_1$, $c_2=\delta$ and $c_3=-\delta$.
The first two of these relations are true within the uncertainty of the fit parameters
at each individual $N$ at any pair of mass values. The third, $c_2= -c_3$, is less clear simply
because $c_3$ is so poorly determined. It is still true within the large uncertainties.

Note also that while, in principle all the dimensionful parameters in the $1/N$ parametrization have
a ``typical QCD size,'' they do show a hierarchy in that $m_0$ is larger than the coefficient of $J(J+1)/N$,
which is larger than the coefficient of $I(I+1)/N$. This hierarchy means that the spin-dependent and
isospin-dependent mass splittings are small across the multiplet, not just at small $J$ and/or $I$.

Before leaving this section, it might be worthwhile to comment on some details of the fits. Generally,
the part of the spectrum which shows the most deviation is the analog of the $\Sigma-\Lambda$ splitting 
at the lowest $J$ values. Coincidentally, these are the smallest splittings, and hence the most susceptible to
numerics (from the simulation point of view) and higher order corrections (from the point of view of the 
mass formula). 

\subsection{Tentative baryon masses in the large-$N$ limit\label{sec:tentative}}

With the data we have in hand, it is tempting to make an extrapolation to large $N$
and show the ingredients which would be needed to predict the mass of any baryon (with any quantum numbers)
at any $N$. This procedure will be incomplete, of course, because of the low quality of the lattice data sets
and (more importantly) because the range of quark masses over which I have three-flavor data is restricted.
I did not collect three-flavor at lower nonstrange quark masses because the signals become noisy.
(This means that I will not discuss chiral extrapolations, but see Ref.~\cite{Cordon:2013}.)
 I will work in terms of the $SU(2)\times U(1)$ mass formula, or of the color hyperfine mass formulas,
since they have the most direct connection to the input nonstrange and strange quark masses.
In these formulas, all input parameters depend on the nonstrange quark mass. I will make the assumption that
any explicit dependence on the strange quark mass can be included as a linear variation in the nonstrange -
strange
mass difference, which can be parametrized by $\delta m$ as described above. Then, for example, the
$c1$ parameter in the $SU(2)\times U(1)$ mass formula  could be written as
\bee
c_1(m_0,\delta m, N) = c_{10}(m_0,N) + c_{11}(m_0,N)\delta m.
\label{eq:c1}
\ee
In principle, the parameters on the right hand side of Eq.~\ref{eq:c1} have an expansion in powers of
$1/N$,
\bee
c_{10}(m_0,N) = c_{100}(m_0) + c_{101}(m_0)/N. \qquad c_{11}(m_0,N) = c_{110}(m_0) + c_{111}(m_0)/N.
\label{eq:c11}
\ee
I have truncated  these expressions at first order in $1/N$ and $\delta m$ because that is about all I can do
given the quality of my data. Then the large $N$ parametrization involves dimensionful parameters,
like $c_1$ or $\delta m$), or are dimensionless numbers, like $c_{11}(m_0,N)$.
 I will quote the dimensionful ones in units of $1/r_1$ 
the Sommer parameter (recall that $1/r_1= 635$ MeV).
Then the way to use my results would be: Pick  values of the pseudoscalar to vector meson mass ratio,
as a way of specifying the masses of the isodoublet of light quarks and of the strange quark. Look
on the figures which follow and interpolate to the desired nonstrange pseudoscalar to vector meson mass ratio.
Multiply the dimensionless parameters by $\delta m$ and evaluate the mass formula.

It is easiest to begin with the color hyperfine formula. The parameters $m_0$ and $C$ were discussed previously,
as was $B(m_0)= B_0 + B_1/N$.

\begin{figure}
\begin{center}
\includegraphics[width=\textwidth,clip]{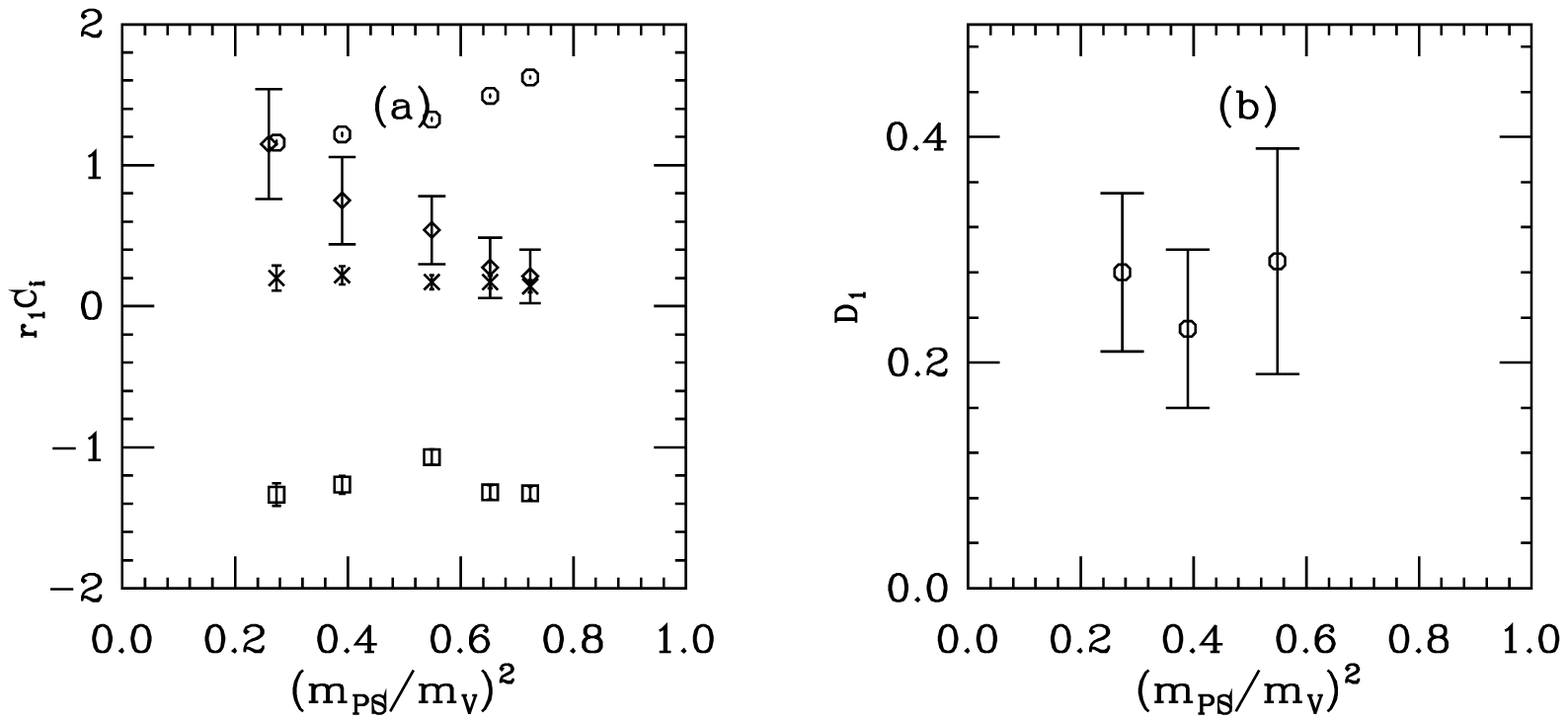}
\end{center}
\caption{ Parameters for the large $N$ limit of the color hyperfine Hamiltonian, as a function of
the squared pseudoscalar-vector mass ratio.
Panel (a) shows the dimensionful parameters $r_1 m_0$ (octagons), $r_1 C$ (squares), $r_1 B_0$
(crosses) and $r_1 B_1$ (diamonds).
Panel B shows the dimensionless parameter $D_1$ of Eq.~{\protect{\ref{eq:d1}}}.
\label{fig:tentcs}}
\end{figure}

The $\delta$ parameter is expected to vary linearly with $\delta m$
\bee
r_1\delta = D_1(m_0) r_1 \delta m,
\label{eq:d1}
\ee
with a coefficient which depends on $m_0$. No $N$ dependence could be seen
in the data, so I extracted $D$ from a linear fit to the data ($r_1\delta = r_1 D_0 +  D_1 r_1 \delta m$)
 at all $N$'s for each matched set. In all cases the best fit $r_1 D_0$ was zero within uncertainties.
These parameters are displayed in Fig.~\ref{fig:tentcs}.

\begin{figure}
\begin{center}
\includegraphics[width=\textwidth,clip]{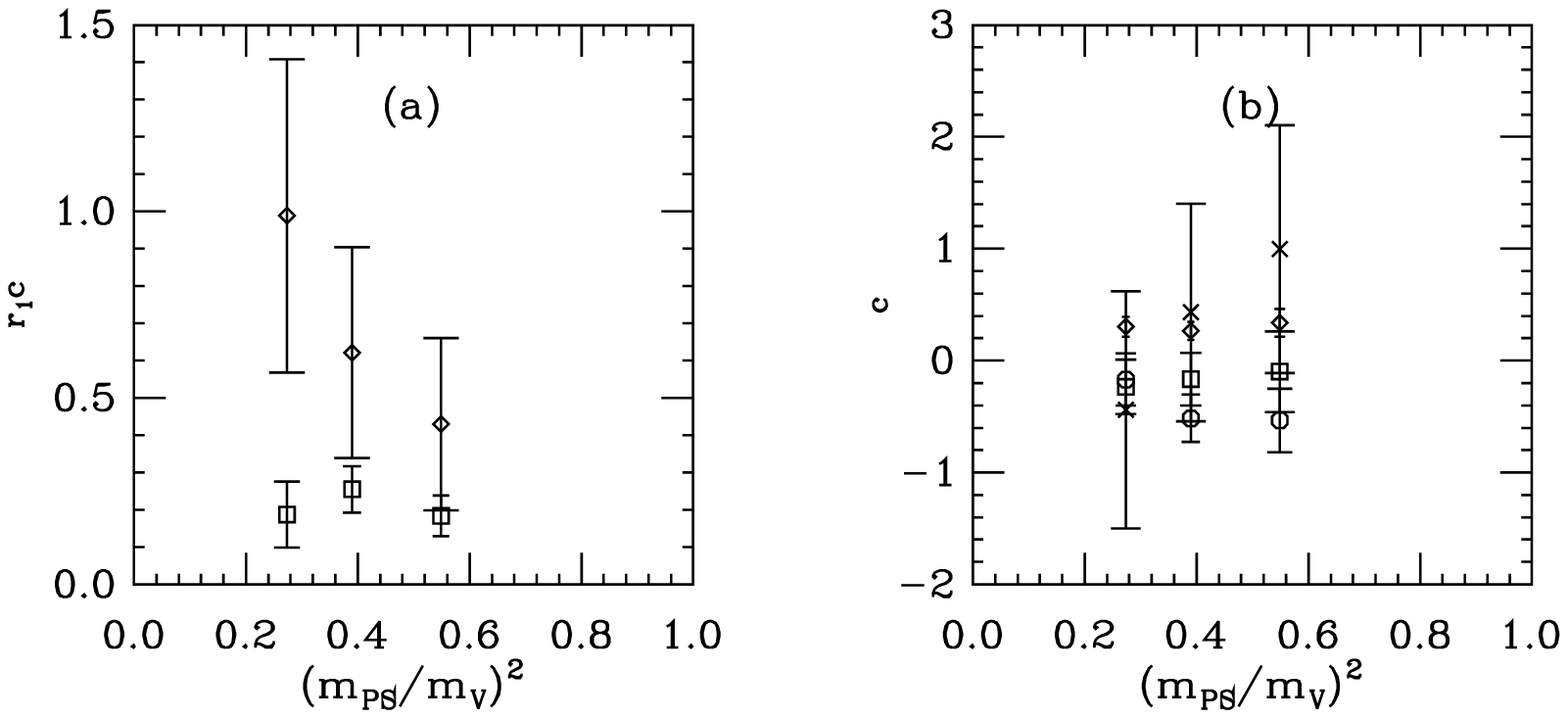}
\end{center}
\caption{ Parameters for the large $N$ limit of the $SU(2)\times U(1)$ Hamiltonian, as a function of
the squared pseudoscalar-vector mass ratio.
Panel (a) shows the dimensionful parameters $r_1 c_{100}$ (squares) and $r_1 c_{101}$ (diamonds),
where $c_{10} = r_1 c_{100} + r_1 c_{101}/N$.
(crosses) and $r_1 B_1$ (diamonds).
Panel B shows the dimensionless parameters $c_{110}$ and $c_{111}$ as octagons and squares; recall
$c_{11}=c_{110}+c_{111}/N$. It also shows $c_{21}$ and $c_{31}$, the slopes with mass of $c_2$ and $c_3$,
as diamonds and squares.
\label{fig:tent7}}
\end{figure}

In the $SU(2)\times U(1)$ formula, we have the three parameters $c_1$, $c_2$ and $c_3$. Like $B$,
$c_1$ has an observable $N$ dependence with, in all, two dimensionful and two dimensionless parameters
to be given.
A fit shows that $c_2$ and $c_3$ are linear in $\delta m$. No discernible $N$ dependence survives
the fit, and so there are two more dimensionless parameters to display. 
These parameters are displayed in Fig.~\ref{fig:tent7}.

\section{Conclusions\label{sec:conclusions}}

Generally, the qualitative expectations of 
either the simple color hyperfine model, or either of the more general $1/N$ mass Hamiltonians successfully reproduce
all my data. The coefficients track with the difference in strange versus nonstrange mass
as expected. Comparisons of different $N$'s at matched parameter values reveal nonleading in $1/N$
behavior for some of the
parameters, most notably the coefficient of $J(J+1)$. 

I am not really sure how good a job of computing spectroscopy one has to do, for systems which do not exist in the real world.
Nevertheless, let us ask what it would take, to remove the word ``tentative'' from the title of Subsec.~\ref{sec:tentative}.
First, because figures like Figs.~\ref{fig:m0vs1n} and \ref{fig:bvs1n} show curvature in the variation of fit parameters
with $1/N$,
 a real extrapolation of the terms in the mass Hamiltonians to $1/N\rightarrow 0$
 requires at least one more value of $N$ so that a higher order polynomial fit in $1/N$ will have a nonzero number of degrees of freedom.

Next,  while all published studies of fermionic QCD I know of use the quenched approximation,
I believe that future work ought to be done with dynamical fermions rather than in quenched approximation.
The large-$N$ limit of QCD shares many features of the large-$N$ limit, but the approximation and the limit really do not commute.
When I tried to push to small quark masses, I encountered exceptional configurations, which are
quenching artifacts. And, any deeper analysis of the data
probably takes us into chiral extrapolations, which are simply different for quenched QCD than for unquenched QCD.
Such simulations may not be a completely daunting task, at least for moderate quark masses.

It might be worth remarking that there are many different large-$N$ limits discussed in the literature.
For example, the quarks could be put into the two-index antisymmetric
 representation of the gauge group
\cite{Armoni:2003fb,Armoni:2003gp,Armoni:2004uu,Cherman:2012eg}.
For $N=3$, this representation is equivalent to the conjugate of the fundamental representation.
To study any of these different  large-$N$ limits in lattice simulations probably requires only human
 persistence
(associated with writing the appropriate code to build states and calculate the appropriate correlators),
and probably only small computer resources, at least for quenched pilot projects.

And to conclude with one sentence, $1/N$ regularities are present in all the $N$'s I studied, and they were
very easy to see.

\begin{acknowledgments}
I thank 
R.~Lebed
for discussions about this subject, and for carefully reading a draft of the manuscript.
I am grateful for the encouragement of A. Hasenfratz, to look for minus signs.
The conversion of the MILC code to arbitrary number of colors was done with
Y.~Shamir and B.~Svetitsky.
This work was supported in part by the U.~S. Department of Energy.
Computations were performed on the University of Colorado theory group's cluster.

\end{acknowledgments}


\end{document}